\documentclass[12pt]{article}
\usepackage{amsmath,amsfonts,epsfig}
\numberwithin{equation}{section}

\ifx\ltimes\undefined
\newcommand{\ltimes}{{\kern3pt\hbox{\vrule width 0.4pt height 5.30pt
depth .0pt}\kern-1.76pt\times\kern1pt}} \fi

\def\bid{\hbox{1\hspace{-0.04in}I}} %blackboard bold 1

\def\Z {\mathbb{Z}}
\def\R {\mathbb{R}}

\def\ti{\tilde}
\def\pa{\partial}
\def\a{\alpha}
\def\b{\beta}

\def\d{\delta}
\def\g{\gamma}

\def\l{\lambda}
\def\m{\mu}
\def\n{\nu}
  
\def\r{\rho}                                     %     \varrho
\def\s{\sigma}                                   %     \varsigma
\def\si{\sigma}                                   %     \varsigma

\def\G{\Gamma}

\def\cG{{\cal G}}
\def\cX{{\cal X}}

\begin{document}

\begin{titlepage}
\begin{flushleft}
\hfill  Imperial/TP/050201 \\
\hfill  QMUL-PH-05-05\\
\hfill  hep-th/0503114 \\

\end{flushleft}
\vspace*{8mm}

\begin{center}

{\Large {Flux Compactifications of String Theory\\ on Twisted
Tori}} \\

\vspace*{12mm}

{ C.M.~Hull$^1$ and R.A.~Reid-Edwards$^{1,2}$
} \\
\vspace*{7mm}

{\em $^1$The Blackett Laboratory, Imperial College London} \\
{\em Prince Consort Road, London SW7 2AZ, U.K.} \\

\vspace*{4mm}

{\em $^2$Department of Physics, Queen Mary, University of London} \\
{\em Mile End Road, London E1 4NS, U.K.} \\

\vspace*{12mm}

\end{center}

\begin{abstract}

Global aspects of Scherk-Schwarz dimensional reduction are
discussed and it is shown that it can usually be viewed as arising
from a  compactification on the compact space obtained by
identifying a (possibly non-compact) group manifold ${\cal G}$
under a discrete subgroup $\Gamma$, followed by a truncation. This
allows a generalisation of Scherk-Schwarz reductions to string
theory or M-theory as   compactifications on
 ${\cal G}/\Gamma$, but only in those cases in which there is a
 suitable discrete subgroup of ${\cal G}$. We analyse such
 compactifications with flux and investigate the gauge symmetry and its
 spontaneous breaking. We discuss the   covariance under $O(d,d)$, where $d$ is the dimension of the group ${\cal G}$, and the relation to reductions with duality twists.
 The compactified theories promote a subgroup of the $O(d,d) $ that would arise from a toroidal reduction to a gauge symmetry, and we discuss the interplay between the gauge symmetry and the $O(d,d,\Z)$ T-duality group, suggesting the role that T-duality should play in such compactifications.

\end{abstract}

\vfill

\noindent {Email: { c.hull@imperial.ac.uk}, {
r.reid-edwards@imperial.ac.uk} }

\end{titlepage}

\newpage

\section{Introduction}

In \cite{Scherk:1979zr}, Scherk and Schwarz proposed
two related forms of dimensional reduction of field theories,
 both of which led to
non-abelian gauge symmetries, a scalar potential and mass terms.
Somewhat confusingly, both have been referred to as Scherk-Schwarz
reductions in the literature. In one type, a
theory with a global duality symmetry is reduced on a circle or
torus with a duality twist or monodromy around each circle. Following
\cite{Dabholkar:2002sy}, we will refer to these as
reductions with a duality twist.

In the other type of reduction  introduced in
\cite{Scherk:1979zr},
 the
dependence of   fields on the internal coordinates $y^i$ is
through a matrix $\s ^i_m (y)$, so that for example the internal
components of the metric $g_{ij}(x,y)$ lead to scalar fields
$\phi_{mn}$ depending only on the remaining external coordinates
$x$ through the ansatz
\begin{equation}\label{metric ansatz}
g_{ij}(x,y)=\phi_{mn}(x)\s^m_i (y)\s ^n_j(y)
\end{equation}
where $\s^m_i(y)$ is the inverse of  $\s ^i_m (y)$. This leads to
a reduced theory in which the $y$-dependence drops out completely
provided the matrices $\s ^i_m (y)$ satisfy the constraint that
the coefficients
\begin{equation}
f^m_{np}=-\s^i_n\s ^j_p(\pa _i \s _j^m -\pa _j \s _i^m)
\end{equation}
are constant. Then the one-forms $\sigma^m= \sigma _i^m (y) dy^i$
satisfy  the structure equation
\begin{equation}\label{maurer-cartan eqn}
d\sigma^m+\frac{1}{2}f^m_{np}\sigma^n \wedge \sigma^p=0
\end{equation}
 and the
integrability condition for this is that the constants $f^m_{np}$
satisfy the Jacobi identity and so are the structure constants for
a Lie group ${\cal G}$. In (\ref{metric ansatz}), the ansatz
$g_{ij}(x,y)=\phi_{ij}(x)$ that would be used for a toroidal
reduction is \lq twisted' by the matrices
 $\s ^i_m (y)$ and so the reduction is sometimes referred
 to as reduction on a \lq twisted torus', and  we will use this
 terminology here. However, we will be particularly interested in the
 global structure and we will see that although the internal space looks
 like a torus locally, the global structure can be quite different, and
 so this terminology can be rather misleading.
 In many standard cases,
 such as those discussed in section 3, the internal space is in fact a torus bundle over a circle or torus, so that the name is appropriate, but other examples include those in which the internal space is a compact group manifold, so that the internal space is very different from a torus, twisted or otherwise.

We will be interested here in flux compactifications in which
the $(p+1)$-form field strength
$\widehat{G}$ for a $p$-form gauge field has a flux of the form
\begin{equation}
\frac{1}{(p+1)!} K_{mn...p+1} \s^m\wedge \s^n\wedge
....\wedge\s^{p+1}+\dots
\end{equation}
where $K_{m_1 m_1...m_{p+1}}$ are constant coefficients, satisfying constraints   that ensure that this form is closed.

In \cite{Scherk:1979zr},   reductions with duality twists arose as
particular examples of reductions on   twisted tori, but this is
not true in general, as will be discussed below. The reductions of
\cite{Scherk:1979zr} are  of field theories such as
supergravities, with a truncation to a lower-dimensional field
theory that is independent of the extra coordinates $y$. An
important feature is that both kinds of Scherk-Schwarz reductions
allow consistent truncations
\cite{Duff:1984hn,Duff:hr,Cvetic:2000dm,Cvetic:2003jy}, in the
sense that it is consistent with the full higher-dimensional field
equations to set all the massive Kaluza-Klein modes to zero while
keeping a finite number of light or massless fields non-zero (this
is not true for generic compactifications; for example, Calabi-Yau
compactifications are not consistent in this sense).

A key question is whether such reductions can be extended to the
full Kaluza-Klein theory or string theory in a way that gives
sensible lower-dimensional physics with a mass-gap. This is a
non-trivial question as there are many cases where such extensions
do not work. For example, there are gauged supergravities with
non-compact gauge group whose lift to higher dimensions is to a
background with non-compact \lq internal' space
\cite{Hull:1988jw}. In such cases, there can be a consistent
truncation to a lower-dimensional supergravity, but if the
Kaluza-Klein spectrum of modes depending on the internal
coordinates are included, one finds a continuous spectrum without
mass gap, so that the theory cannot be properly regarded as a
lower-dimensional theory at all, but is best interpreted in the
full higher-dimensional space-time
\cite{Nicolai:1984jg,Gibbons:2001wy,Hull:2001ii}.
 As the gauge groups arising from Scherk-Schwarz
reductions are typically non-compact, there is a danger that the
full lift of a Scherk-Schwarz reduction   could be to such a \lq
non-compactification'. If a reduction scheme can be regarded as a
compactification on a compact internal space, then it can be
extended to the full  Kaluza-Klein theory or string theory with a
mass gap. It is to this question of whether Scherk-Schwarz
reductions can be viewed as reductions on compact spaces that we
now turn.

Consider first the reduction  on the twisted torus. The simplest
way of realising this is if the internal manifold is the group
manifold of ${\cal G}$ with $\s^m $ the left-invariant
Maurer-Cartan forms, which automatically satisfy
(\ref{maurer-cartan eqn}). For compact groups, this of course
leads to a compactification, although the reduction ansatz is not the usual one. For a non-compact group, the internal
group manifold is non-compact,  but one can still consistently truncate to
the light sector that is independent of the internal
group-manifold coordinates, recovering the lower-dimensional field
theory of the Scherk-Schwarz reduction. However, in this case
there would be no mass-gap and so no satisfactory way of extending
to the full theory.
  As was pointed out in \cite{Scherk:1979zr}, the group ${\cal G}$ being non-compact does
not necessarily imply that   the internal manifold is non-compact,
so that it is possible for the internal space to be compact so
that there is a mass-gap and a well-defined Kaluza-Klein
reduction. For non-compact groups, if the  internal space is
compact, then it cannot be the group manifold. Nonetheless, for
the reduction to be well-defined, the matrices $\s ^i_m (y)$
should exist globally on the internal manifold, so that there are
globally defined one-forms $\s^m$. This implies that the internal
manifold is parallelisable, and so locally  must be a group
manifold. Thus the internal manifold must be a group manifold
identified under the action of  some freely acting discrete group
$\G$. The group manifold admits a natural left action $\cG _L$ and
a right action $\cG _R$, but only the left action preserves the
one-forms $\s^m$ appearing in the ansatz, so the discrete group
$\G$ must be a subgroup of $\cG _L$. Then the internal space must
be of the form $\cG /\G$ for some discrete left-acting $\G \subset
\cG_L$, and we are particularly interested in the cases in which
$\G$ can be chosen so that $\cG /\G$ is compact. Such a $\G$ is
said to be cocompact, and not all groups have cocompact discrete
subgroups. Groups without a cocompact discrete subgroup give
Scherk-Schwarz reductions of supergravity that cannot be extended
to compactifications of  string theory in this way.
 Note that in general $\G$ need not be unique, and we can also consider discrete quotients $\cG /\G$ for compact $\cG  $.

One of the aims of this paper is to study   reductions on twisted
tori
 from the global viewpoint, showing that they can be regarded
as dimensional reductions on compact  internal spaces of the form
$\cG /\G$ (when $\G$ can be chosen so that $\cG /\G$ is compact)
and so can be extended to string theory. There has been much
interest in applying Scherk-Schwarz-type reductions to
supergravity (including [2,11-25])  and to string theory
(including
\cite{Dabholkar:2002sy,Hull:2003kr,Hull:1998vy,Kachru:2002sk,deBoer:2001px,Kachru:2002he}),
but there are important issues as to how to properly define the
full string theory (as opposed to its supergravity limit).
Regarding the reduction as a compactification on $\cG /\G$ allows
a proper definition of the full string theory. It is important
that the ansatz for the metric, gauge fields and fluxes of the
background is invariant under rigid $\cG_L$, so that one can
identify under a discrete subgroup of $\cG_L$.

Identifying the theory under the action of a discrete
subgroup of the isometry group will break part of the isometry
symmetry, and so affect the gauge symmetry of the reduced theory.
There has been some confusion in the literature as to  the way
that gauge symmetry works in these compactifications, and our
viewpoint clarifies some of  the issues involved. We   give a
careful treatment of the gauge symmetry, and its breaking.

The simplest case is that in which the internal space is the group
manifold for a compact group $\cG$. The conventional Kaluza-Klein
ansatz for compactification on the group manifold introduces gauge
fields for $\cG_L\times \cG_R$ and has $\cG_L\times \cG_R$ local
gauge symmetry. The ground state in which the internal metric is
the bi-invariant Killing metric $\d_{mn}\s^m_i (y)\s ^n_j(y)$ has
isometry $\cG_L\times \cG_R$ and so preserves the full gauge
group. The Scherk-Schwarz ansatz for the same compact group
manifold is a truncation of this to the sector invariant under
$\cG_L$ and only has gauge fields for $\cG_R$. This ansatz is
invariant under local $\cG_R$ transformations but only under rigid
$\cG_L$ transformations, and the Killing metric gives a vacuum
preserving all these symmetries. The conventional $\cG_L\times \cG_R $
ansatz does not
allow a consistent truncation, but the Scherk-Schwarz one does.
Since only invariance under $\cG_L$  is required, a more general
ansatz for the ground state metric with $g_{ij}=\phi_{mn}\s^m_i
(y)\s ^n_j(y) $ for any positive definite matrix $\phi_{mn}$ is
permissable, and this
 breaks $\cG_R$ to the subgroup preserving
this   background value for the  scalars $\phi_{mn}$.

Consider now the case of non-compact $\cG$. In order to be able to
factor by a discrete subgroup of $\cG_L$, one must use the
Scherk-Schwarz ansatz which is invariant under rigid $\cG_L$
symmetry, and the resulting theory
  has $\cG_R$ gauge symmetry, but the identification under $\G$
   breaks the $\cG_L$ symmetry. The space $\cG/\G$
still has an action of $\cG_R$, with infinitesimal transformations
given by well-defined vector fields on $\cG/\G$,
 and introducing gauge fields in the adjoint of
$\cG$ gives a theory with local $\cG_R$ symmetry. However, this
non-compact gauge symmetry is necessarily non-linearly realised
and so is broken to a  linearly realised compact subgroup. (A
non-compact symmetry can have no unitary representation on a
finite number of fields, and so must be non-linearly realised.)
The scalar fields $\phi_{mn}$ define a positive definite metric on
the internal space, and the
 expectation value of this for any ground state will not in general be invariant under
 the full non-compact $\cG_R$ symmetry but will spontaneously break this to a subgroup.

We turn now to consider reductions with duality twists, and their
implementation in string theory. A theory in $D+1$ dimensions with
a global symmetry $G$ is reduced on a circle with the dependence
of   fields $\Psi (x,y)$ (in some representation of $G$)  on the
circle coordinate $y$ given by a $y$-dependent $G$ transformation
$\Psi (x,y)=\exp (My) \psi(x) $ where $M$ is a generator of $G$ in
the appropriate representation. However, in the full theory in which all massive modes are kept, $G$
is typically broken to a discrete subgroup $G(\Z)$.

 For the reduction with duality twist,  on going round the circle, $y \to y +2\pi$   there is a monodromy
 ${\cal M}=\exp (2\pi M)$. In order for $\Psi (x,y+2\pi)={\cal M}\Psi (x,y) $ to be well-defined,
 the monodromy ${\cal M}$ must be in the symmetry group $G(\Z)$ \cite{Hull:1994ys}.
 However, if only $G(\Z)$ is a symmetry of the full theory, it is not immediately clear how to
 extend the continuous action of $G$ that is needed in the ansatz from the low-energy  sector to the full theory with massive modes and only $G(\Z)$ symmetry,
 and care is needed in the full definition of the theory.
 For the case of the toroidal reduction on $T^d$ followed by a reduction on a circle with a $Gl(d,\Z)$ twist, this was resolved in \cite{Hull:1998vy}. There it was shown that the reduction
 can be viewed as a
reduction on a $d+1$-dimensional compact space that is a $T^d$
bundle over a circle. As this is a compactification, it can be
used in string theory.
Moreover, this compact space is locally a group manifold for a non-semi-simple lie group, as we shall show in section 3.
 Then  the torus bundle is
locally the group manifold for a group ${\cal G} $, and globally
is of the form ${\cal G}/\G$ for a  discrete subgroup $\G$, and so
this is a reduction on a twisted torus. In this case, the twisted
torus is a torus bundle over a circle, so that  in this case it is
actually a topological twisting of a torus.

In this example, the twist is by a duality group that has a
geometric realisation in the higher dimensional space -- in this
case $GL(\Z)$ is the group of large diffeomorphisms of $T^d$.
However, the T-duality or U-duality groups that arise in string
theory have many elements that are not geometric in this sense,
and such non-geometric twists have been considered in
\cite{Dabholkar:2002sy,Flournoy:2004vn}.
 In such cases, there is not in general any way of
realising such reductions as reductions on a geometric background,
and these cannot be realised as reductions on twisted tori, and
any such lift would lead to what might be called a non-geometric
background; such backgrounds have been discussed in
\cite{Hull:2004in,Hellerman:2002ax}. For such duality twist
reductions corresponding to non-geometric backgrounds, one needs
to check that these really are consistent string backgrounds.
However, in some cases, these can be realised as \lq
compactifications' of F-theory (or one of its generalisations) on
a twisted torus \cite{Hull:1998vy}. Here, we will restrict
ourselves to geometric backgrounds that can be realised as twisted
tori.

In this paper, we will consider generalised   Scherk-Schwarz
compactifications on $d$-dimensional  twisted tori ${\cal G}/\G$
with flux for the 3-form field strength $H$ (the NS-NS 3-form in
type II or heterotic superstring theory); other fluxes will be considered elsewhere.
The flux will be taken to be  of the form
\begin{equation}
K= \frac{1}{6}K_{mnp} \s^m\wedge \s^n\wedge \s^p
\end{equation}
for some constant coefficients $K_{mnp}$ and so is manifestly
invariant under $\cG_L$. The gauge group contains the isometry
group ${\cal G}$ with $d$ gauge generators $Z_m$ corresponding to
the Killing vectors of the internal space.
 For string theory,
 there are, as we shall see, an additional $d$ generators  $X^m$ associated with the gauge fields arising from the
 reduction of the 2-form gauge field, and the structure constants of the $2d$ dimensional gauge group generated by $(Z_m,X^m)$ depend on the  3-form flux.
It was shown in \cite{Odd} that the final theory has an elegant
formulation that is covariant under the action of $O(d,d)$, with
the gauge generators $(Z_m,X^m)$ combining to form a vector of
$O(d,d)$, and manifestly invariant under the gauge group ${\cal
G}$, which can be viewed as a subgroup of $O(d,d)$.
Indeed, the low energy field theory can be thought of as arising from a gauging
of a subgroup of the $O(d,d)$ symmetry that would arise in a torus reduction.
For the heterotic string, the $O(d,d)$ symmetry is contained in $O(d,d+16)$
while for type II strings it is contained in the U-duality group.
We review the
 construction of \cite{Odd} for the common sector of
the type II and heterotic theories, focusing on the $O(d,d)$ subgroup
of the full symmetry.

In \cite{Cvetic:2003jy}, the case of compact $\cG$
was considered and there it was found that field redefinitions
simplified the structure of the theory, so that the resulting
gauge algebra $G$ simplified to the   semi-direct product  of
$\cG$ with $U(1)^d$. We find that in the non-compact case, such
redefinitions are not possible in general, and the presence of
flux necessarily leads to non-trivial structure of the gauge group
$G$.

The structure of the paper is a follows. In section 2 we discuss
the global structure of twisted torus reductions, showing that
they are reductions on discrete quotients of group manifolds. In
section 3 we discuss reductions with duality twists and show that
they are twisted torus compactifications when the duality symmetry
is of geometrical origin. In section 4 we discuss dimensional
reduction on twisted tori with fluxes and in section 5   apply
this to string theory, and discuss the gauge symmetry and $O(d,d)$
covariance that were found by Kaloper and Myers \cite{Odd}. In
Section 6 we analyse the breaking of the gauge symmetry. The final
section discusses the generalisations of our results to M-theory
compactifications, and addresses the role of the $O(d,d)$
covariance. On the one hand, a subgroup of the $O(d,d) $ symmetry
of the toroidal reduction of the low-energy field theory has been
promoted to a gauge symmetry, but in the full string theory one
expects the $O(d,d)$ symmetry to be broken to a discrete T-duality
subgroup, so that the issue of whether a subgroup of $O(d,d)$ can
be gauged in the full string theory arises. We resolve the issue
of the relation between the discrete symmetry and the local gauge
symmetry, and discuss the status of T-duality in such
compactifications.

\section{Twisted Tori and Group Manifolds}

In this section we  review Scherk-Schwarz reduction of field
theory and show that in many cases it can be viewed as a compactification on a
 compact twisted torus, and so can be extended to string theory.

\subsection{Scherk-Schwarz Reduction}

We shall consider Scherk-Schwarz dimensional reduction
\cite{Scherk:1979zr} in which the internal space is a
$d$ dimensional
  manifold ${\cal X}$ with
coordinates $y^i$ and a basis of nowhere-vanishing one-forms
$\sigma^m$ specified by a vielbein $\sigma _i{}^m(y)$
\begin{equation} \sigma^m= \sigma _i{}^m (y) dy^i
\end{equation}
In the ansatz of \cite{Scherk:1979zr}, the internal components
$T_{ij...k}$ of a tensor field $T_{MN...P}$ are taken to have $y$
dependence given only by the frame fields
\begin{equation}
T_{ij...k}(x,y)=T_{mn...p}(x)\si _i{}^m\si _j{}^n...\si _k{}^p
\end{equation}
defining scalar fields $T_{mn...p}(x)$ in the reduced theory, so
that for example the internal metric takes the form (\ref{metric
ansatz}). The frame fields satisfy the structure equation
\begin{equation}
\label{eq:maurer-cartan} d\sigma^m+\frac{1}{2}f^m_{np}\sigma^n
\wedge \sigma^p=0
\end{equation}
where
the coefficients $f^m_{np}$
are the structure constants for some Lie group
${\cal G}$, satisfying  the Jacobi identity
\begin{equation}
\label{eq:ff=0} f_{[mn}^qf_{p]q}^t=0
\end{equation}
Such manifolds ${\cal X}$ are sometimes referred to as
twisted tori in the cases in which the coordinates $y^i$ satisfy
periodicity conditions, and the matrix $ \sigma _i{}^m (y)$ can be
thought of as defining the twisting of the frames with respect to
the coordinate basis. The structure equation implies that  ${\cal
X}$ is locally isomorphic to the group manifold ${\cal G}$, but
this need not be true globally; in general, ${\cal X}={\cal
G}/\Gamma$ where $\Gamma$ is a discrete subgroup of ${\cal G}$,
and ${\cal X}$ can be compact even if ${\cal G}$ is non-compact.
Then Scherk-Schwarz reduction can be viewed as compactification on a group
manifold or
 ${\cal X}={\cal
G}/\Gamma$ (if this is compact).
Note that in general $ {\cal
G}/\Gamma$ will not be a group manifold.

\subsection{Compact Groups}

The simplest case is that in which ${\cal X}$  is the group
manifold for a compact group $\cal G$, with $\sigma^m$ the
left-invariant Maurer-Cartan one-forms, so that if $t_m$ are Lie
algebra generators $\sigma ^mt_m= g^{-1}dg$ for some $g\in \cG$.
It will be useful to introduce the Cartan-Killing metric for
${\cal G}$ given by $\eta_{mn}=\frac{1}{2}f_{mp}^qf^p_{nq}$ (which
   is proportional to $\d_{mn}$ for compact $\cal
G$). Then the metric
\begin{equation}
\label{Invariant metric} ds^2=\eta_{mn} \s^m\s^n
\end{equation}
is invariant under the isometry group $\cG_L\times \cG_R$, where
$\cG_L$ is the left action $g \to k_Lg$ and $\cG_R$ is the right
action $g \to gk_R$ for $k_L,k_R \in \cG$.
 The
inverse vielbein $\s_m^i$ can be used to define the left-invariant
vector fields
\begin{equation}
Z_m=\s_m^i \frac{\pa}{\pa y^i}
\end{equation}
with Lie bracket
\begin{equation}
[Z_m,Z_n]= f^p_{mn}Z_p
\end{equation}
There are also right-invariant one-forms $\ti \s^m$ with $\ti
\sigma ^mt_m=dgg^{-1} $ satisfying
\begin{equation}
d\ti \sigma^m-\frac{1}{2}f^m_{np}\ti\sigma^n \wedge \ti\sigma^p=0
\end{equation}
and right-invariant vector fields $\ti Z_m$ given by
\begin{equation}
\ti Z_m=\ti \s_m^i \frac{\pa}{\pa y^i}
\end{equation}
The Killing vectors $Z_m$ generate   $\cG_R$ which leaves $\ti
\s^m$ invariant while the Killing vectors $\ti Z_m$ generate
$\cG_L$ which leaves $\s^m$ invariant.

A conventional Kaluza-Klein reduction on a group manifold would
take the bi-invariant metric (\ref{Invariant metric}) as the
vacuum  and would introduce Kaluza-Klein gauge fields for the full
isometry group $\cG_L\times \cG_R$. The gauge group for the
dimensionally reduced theory would then be $\cG_L\times \cG_R$
with vector fields $A^m_\mu$ for $\cG_R$ and $\ti A^m_\mu$ for
$\cG_L$. The coordinates of the spacetime are $\{x^{\mu},y^i\}$
where $x^{\mu}$ are the coordinates of the uncompactified part of
spacetime. However, for generic theories there is no consistent
truncation to a dimensionally reduced theory with a finite number
of fields and it is necessary to keep the full Kaluza-Klein tower
of states \cite{Duff:hr}. This can be easily seen from the
 Einstein equations for the metric $g_{\m\n}$,
the right-hand-side of which includes a term $(Z_m \cdot \ti Z_n)
F^m_{\m\r} \ti F^{n\r}_\n$ where $F^m_{\mu\nu}$ and
$\tilde{F}^m_{\mu\nu}$ are the field strengths of $A^m_{\mu}$ and
$\tilde{A}^m_{\mu}$ respectively. The fact that $(Z_m \cdot \ti
Z_n) $ depends non-trivially on $y$ means that $g_{\m\n}$ must be
taken to be a function of both $x$ and $y$, and a truncation to a
finite set of fields depending only on $x$ would be inconsistent
with the field equations; it is necessary to keep all the massive
Kaluza-Klein modes $g_{\m\n}^N(x)$ from the expansion of
$g_{\m\n}(x,y)$.
 However, for the theories with a metric, 2-form and dilaton that arise in string theory, there is evidence that there is a consistent truncation to a theory with gauge group $\cG_L\times \cG_R$
 \cite{Cvetic:2000dm,Cvetic:2003jy}.

 In the Scherk-Schwarz reduction, by contrast, the ansatz introduces only
 Kaluza-Klein gauge fields $A^m_\mu$  for the isometry group   $\cG_R$, and truncates the full Kaluza-Klein spectrum to the sector of singlets under $\cG_L$.
 This does allow a consistent truncation to a finite number of fields, as the Einstein equations now  only involve source terms such as
 $(Z_m \cdot  Z_n) F^m_{\m\r}  F^{n\r}_\n$
 and the fact that $(Z_m \cdot  Z_n) =\eta_{mn}$ is independent of $y$
 allows a truncation to a metric $g_{\m\n}(x)$ that is independent of $y$.
 The ansatz for the metric is
 \begin{equation}
 \label{invariant metric 2}
ds^2_D=e^{2\alpha\varphi} g_{\mu \nu}dx^\m dx^\n
+e^{2\beta\varphi}g_{mn}\nu^m\nu^n
\end{equation}
where the one-forms $\nu^m$ are
\begin{equation}
\nu^m=\sigma^m - A^m
\end{equation}
and $\a,\b$ are constants. Dimensional reduction gives rise to a
metric $g_{\mu \nu}(x)$, $d$ Kaluza-Klein one-form gauge fields
$A^m_\mu(x)$, and $d(d+1)/2$ scalars  $\varphi(x)$ and
$g_{mn}(x)$, where $g_{mn}(x)$ is a positive definite symmetric
matrix with unit determinant. This ansatz is invariant under rigid
$\cG_L$ transformations, and under local $\cG_R$ transformations
in which the parameters depend on $x^\m$ and the $A^m$ transform
as gauge fields, while the scalar fields $g_{mn}(x)$ transform as
the bi-adjoint. A vacuum in which the scalars have the expectation
value $\bar g_{mn}=\eta_{mn}$ will be invariant under $\cG_R$
while any other expectation value $\bar g_{mn}$ will break the
gauge symmetry to the subgroup preserving $\bar g_{mn}$.

The ansatz for antisymmetric tensor gauge fields is the most
general one that is invariant under $\cG_L$, so that for a
$p$-form potential
\begin{equation}
\label{B-field ansatz}
\widehat{B}_{(p)}=B_{(p)}+B_{(p-1)m}\wedge\nu^m+\frac{1}{2!}B_{(p-2)mn}\wedge
\nu^m\wedge\nu^n+...+\frac{1}{p!}B_{(0)m_1 m_2
m_3...m_p}\nu^{m_1}\wedge...\wedge\nu^{m_p}+\varpi_{(p)}
\end{equation}
where $B_{(p)}$ is an $p$-form gauge field on $M_d$ and a flux
term $\varpi_{(p)} $ is included (see section 3.2). Again, the
invariance under $\cG_L$ guarantees consistency
\cite{Duff:1984hn}. We will be particularly interested here in the
case in which $p=2$, in which case the reduction gives a 2-form,
$d$ vector fields $B_{(1)m}$ and scalars $B_{(0)m_1 m_2}$. The
reduction of   antisymmetric tensors, and the constraints that
must be imposed on the flux, will be discussed further in section
 4.2.

\subsection{Non-Compact Groups}

Consider now the case of non-compact $\cG$, so that $\eta_{mn}$ is
no longer positive definite. If the group is non-semi-simple,
$\eta_{mn}$ will be non-invertible. For a spacetime which is a
(possibly warped)  product   of a spacetime
$M$ and the non-compact group manifold $\cG$, then one can consider
$\cG$ as an internal space and attempt to expand in terms of modes
on $\cG$, but the resulting theory has no mass-gap in general and
cannot be properly regarded as a $d$-dimensional theory.
Nonetheless, there is still a consistent truncation   as above to
a finite set of fields in $M$, using the same ansatz
(\ref{invariant metric 2}). In this case, there is usually no
$\cG_R$-invariant ground state. This is because the internal
metric  $g_{mn}$ of the ground state is required to be
positive-definite and usually there is no positive definite
invariant metric for a non-compact group. The Cartan-Killing
metric is invariant but not positive-definite. For semi-simple
non-compact groups, the only invariant metric is the
Cartan-Killing metric, but   non-semi-simple groups sometimes have
positive invariant metrics \cite{Figueroa-O'Farrill:1994ns}. One
cannot set $g_{mn}=\eta_{mn}$, and any expectation value for
$g_{mn}$ will break the gauge symmetry to the subgroup of $\cG_R$
preserving the expectation value $\bar g_{mn}$.

As  the ansatz (\ref{invariant metric 2}),(\ref{B-field ansatz})
is invariant under rigid  ${\cal G}_L$ transformations, one can
identify the internal space under the action of a discrete
subgroup $\Gamma$   of ${\cal G}_L$ so that the internal space is
the left coset ${\cal X}={\cal G}/\Gamma$. If the discrete
subgroup is chosen so that ${\cal X}={\cal G}/\Gamma$ is compact,
then one can perform a compactification with this internal space
and there will be a Kaluza-Klein spectrum with a mass-gap governed
by the size of $\cX$. Then $\cX$ is locally isomorphic to the
group manifold $\cG$, and much of the structure will be the same.
In particular, the left-invariant one-forms $\s^m$ are
well-defined on $\cX$ and satisfy the structure equation
(\ref{eq:maurer-cartan}) so that this is a Scherk-Schwarz
compactification, and any such compactification must be of this
type. The low-energy effective physics in $M$ only depends on the
local structure of $\cX$, and so must contain the consistent
truncation of the theory on $\cG$ with gauge symmetry $\cG_R$
described in the previous paragraph. This must then be a
consistent truncation of the theory on ${\cal G}/\Gamma$ also.

Importantly, not every group $\cG$ has a cocompact discrete subgroup $\G$
  that gives a compact space ${\cal G}/\Gamma$. Only if there is such a $\G$ can the Scherk-Schwarz reduction and truncation of the low-energy field theory be promoted to a Kaluza-Klein compactification or a compactification of the field theory. If there is such a $\G$, it may not be unique and
  compactifications on ${\cal G}/\Gamma$ or ${\cal G}/\Gamma'$ will give
  different theories
 that have
 the same effective low-energy theory.
 In particular, for a compact group $\cG$, we could consider
either compactification on $\cG$ itself, or a quotient ${\cal G}/\Gamma$.

The right-action of $\cG_R$ is well-defined on the left-coset
${\cal G}/\Gamma$, so that the   vector fields $Z_m$ that generate
this action are well-defined on the quotient space. For any given
expectation value   $\bar g_{mn}$ of the internal metric, only a
subset of the $Z_m$ will be Killing vectors, and the gauge
symmetry is spontaneously broken to the compact subgroup of
$\cG_R$ generated by the $Z_m$ which are Killing vectors for $\bar
g_{mn}$. The compactified theory has local gauge symmetry under
the full non-compact gauge group $\cG_R$, even though there are no
vacua with a full set of Killing vectors $Z_m$ generating $\cG_R$,
and this is always broken to a compact subgroup in any solution.

%%%%%%%%

\section{Reductions With Duality Twists}

\subsection{Geometric Twists}

In this section we discuss   reductions with duality twists and show that in a large class of cases they are equivalent to
compactifications on twisted tori.

Consider a $D+d+1$ dimensional field theory coupled to gravity. We
reduce the theory on a $d$-dimensional torus $T^d$, with real
coordinates $z^a\sim z^a+1$ where $a=1,2...d$.
This produces a theory in
$D+1$ dimensions with scalar fields
that include those
 in the coset $GL(d,\R)/SO(d)$ arising from the torus
moduli. Truncating to the $z^a$ independent zero mode sector, this
theory has a global $G$ symmetry that contains the $GL(d,\R)$
arising from diffeomorphisms of the torus, while in the full
Kaluza-Klein theory this is broken to the $GL(d,\Z)$ that acts as
large diffeomorphisms on the $d$-torus. In string theory $G$ is
typically broken to a discrete subgroup $G(\Z)$. For supergravity
theories with sufficient supersymmetry, the full set of scalar
fields typically take values in
 the coset $  G/K$, where $K\subset G$ is the maximal
compact subgroup of $G$.
We
denote the action of $G$ on fields $\psi$ of the reduced theory in
some representation of $G$ as $\psi\rightarrow \gamma[\psi]$.

 We reduce to $D$
dimensions on a further circle with periodic coordinate $y\sim
y+1$, twisting the fields over the circle by an element of
$G$   using the
 ansatz \cite{Hull:1998vy,Hull:2003kr,Hull:2002wg,Dabholkar:2002sy}
 \begin{equation}\label{s-s ansatz}
\psi(x^\m,y)=\gamma_y[\psi(x^\m)]
\end{equation}
where $x^\m$ are the $D$ spacetime coordinates. Consistency of the
reduction requires the reduced theory to be independent of $y$,
which is achieved by choosing the form of $\gamma$ to be
\begin{equation}
\gamma(y)=exp\left(My\right)
\end{equation}
for some mass matrix $M$ in the Lie algebra of $G$. (The masses of
the reduced theory are given by the matrix $M$.) The map
$\gamma(y)$ is not periodic, but has monodromy
$\mathcal{M}(\gamma)=\gamma(0)\gamma(1)^{-1}=e^M$ in $G$.
The physically distinct reductions are classified by the conjugacy class of the monodromy  \cite{Hull:1998vy}.

We now focus on the case in which the monodromy is in the geometrical $GL(d,\R)$ subgroup of $G$.
If it is in fact in $G(\Z)$, then the reduction is equivalent to the compactification on a $T^d$ bundle over a circle, with monodromy $\mathcal{M}$  \cite{Hull:1998vy}. We will see that this compact space is  locally a group manifold, i.e. it is of the form $\cG/\G$.

Let
\begin{equation}
ds^2=H(\tau)_{ab}dz^a dz^b
\end{equation}
 be the metric on the
$d$-torus, depending   on the   moduli
$\tau$, which take values in the coset $GL(d,\R)/SO(d)$.
There is a natural action of $GL(d,\R)$
 on the metric and coordinates $z^a$ in which
\begin{equation}\label{coset symmetries}
H_{ab}\rightarrow (U^t)_a{}^cH_{cd}U^d{}_b    \qquad
z^a\rightarrow (U^{-1})^a{}_b z^b
\end{equation}
where $U^b{}_a\in GL(d,\R)$.
This defines the transformation  $\tau \to \tau '$ of the moduli through
 \begin{equation}\label{moduli symmetries}
H_{ab}( \tau')=
 (U^t)_a{}^cH_{cd}( \tau)U^d{}_b
\end{equation}

In the twisted reduction, we introduce dependence on the circle coordinate $y$
through a $GL(d,\R)$ transformation $U=\gamma(y)$ where $\gamma(y)=exp\left(My\right)$.
This defines the $y$-dependence of
$\tau$ through
\begin{equation}
H(\tau(y))_{ab}=(\gamma (y)^t)_a{}^cH(\tau_0)_{cd}\gamma (y)^d{}_a
\end{equation}
for some arbitrary choice of $\tau _0$.
If the monodromy is in $SL(d,\Z)$, which we now assume, then the
twisted reduction is equivalent to the reduction on a $T^d$ bundle over $S^1$ with metric
\begin{equation} \label{metricbun}
ds_{d+1}^2=dy^2+H(\tau(y))_{ab}dz^adz^b=(\sigma^y)^2+H(\tau_0)_{ab}\sigma^a\sigma^b
\end{equation}
where
\begin{equation} \label{forms}
\sigma^y=dy \qquad \sigma(y)^a=\gamma(y)^a{}_bdz^b
\end{equation}

We now consider the group structure of this space.
The forms (\ref{forms}) are globally defined on the torus bundle, and satisfy
\begin{equation}\label{parellelisability}
d\sigma^a+M^a{}_b\sigma^y\wedge\sigma^b=0
\end{equation}
The space is then parallelisable, and locally looks like a group manifold $\cG$ with Maurer-Cartan forms $\s$ associated with the Lie algebra
\begin{equation}\label{duality twist algebra}
[t_a,t_y]=M_a{}^bt_b, \qquad [t_a,t_b]=0
\end{equation}
This algebra can be represented by the $(d+1)\times(d+1)$ matrices
\begin{equation}
t_y=\left(\begin{array}{cc}-M^a{}_b & 0 \\ 0 & 0
\end{array}\right)  \qquad  t_a=\left(\begin{array}{cc}0 & e_a \\ 0 & 0
\end{array}\right)
\end{equation}
where $e_a$ is the $d$-dimensional column vector with a 1 in the a'th  position
and zeros everywhere else.
Coordinates $y, z^a$ can be introduced for the group manifold, with the group element
 $g=g(y,z^a)\in {\cal G}$
 given by
 \begin{equation}
g=\left(\begin{array}{cc}\gamma^{-1}(y) & z \\ 0 & 1
\end{array}\right)
\end{equation}
 Then the
left-invariant Maurer-Cartan forms are given by
\begin{equation}
g^{-1}dg=\left(\begin{array}{cc}-M^a{}_b\sigma^y & \sigma^a \\
0 & 0
\end{array}\right)=\sigma^mt_m
\end{equation}
in agreement with (\ref{forms}).

The left action of
\begin{equation}
h(\alpha,\beta^a)=\left(\begin{array}{cc} {\g}^{-1}(\alpha) &
\beta \\ 0 & 1 \end{array}\right)
\end{equation}
is
\begin{equation}
g(y,z^a)\to h(\alpha,\beta^a)\cdot g(y,z^a)
\end{equation}
and acts on the coordinates
through
\begin{equation}
y\to y+\alpha  \qquad  z^a\to (e^{-M\alpha})^a{}_b z^b+\beta^a
\end{equation}
The $h(\alpha,\beta^a)$ with $\a ,\b^a \in \Z$ can be written as
\begin{equation}
h(\alpha,\beta^a)=\left(\begin{array}{cc} {\cal M}^{-\alpha} &
\beta \\ 0 & 1 \end{array}\right)
\end{equation}
and  form a discrete subgroup $\G$
and we can identify $\cG$ under the action of $\G$, so that the coordinates
$y, z^a$ are subject to the identifications
\begin{equation}
y\sim y+\alpha  \qquad  z^a\sim ({\cal
M}^{-\alpha})^a{}_bz^b+\beta^a
\end{equation}
for $\a, \b^a \in \Z$.
This in general gives
  a compact space,
  and is the required
  twisted torus construction.

\subsection{Non-Geometric Twists and F-Theory}

In some cases, the discussion of geometric twists can be extended
to non-geometric twists, i.e. twists by duality transformations
that do not arise from higher-dimensional diffeomorphisms.
Consider for example the $SL(2,\Z)$ U-duality of the IIB string
theory \cite{Hull:1994ys}. Reducing from 10 to 9 dimensions on a
circle with monodromy in $SL(2,\Z)$ was investigated in
\cite{Hull:2002wg,Bergshoeff:2002mb,Hull:1998vy,Meessen:1998qm}.
As the $SL(2,\Z)$ symmetry is not geometric,  this cannot be
realised as a compactification  on a twisted torus in the usual way. However, it can
be realised as a \lq compactification' of F-theory on the twisted
torus corresponding to  a $T^2$ bundle over $S^1$ with $SL(2,\Z)$
monodromy \cite{Hull:1998vy}. Many other examples can be thought
of  as compactifications of F-theory \cite{Vafa:1996xn} or its
generalisations \cite{Kumar:1996zx}. For example, the reduction of
M-theory to 7 dimensions, followed by a reduction on a further
circle with a twist by an $SL(5,\Z)$ U-duality transformation can
be viewed as a compactification of the $F'$ theory of
\cite{Kumar:1996zx} on a twisted torus constructed as a $T^5 $
bundle over $S^1$ \cite{Hull:1998vy}.

\subsection{Examples with $SL(2)$ Twists}

We now consider the example of $d=2$ in more detail.
 Reducing from $D+3$ dimensions on    $T^2$, and then on a further circle with an $SL(2,\Z)$ twist is equivalent to reducing on a $T^2$ bundle over $S^1$ \cite{Hull:1998vy} with monodromy ${\cal M}=e^M\in SL(2,\Z)$.
 The $T^2$ has moduli $A,\tau =\tau_1+i\tau_2$ where  $A$ is the area of the torus and $\tau$ its complex
structure modulus, and the metric is
\begin{equation}\label{metric}
H(\tau)=A\frac{1}{\tau_2}\left(\begin{array}{cc} 1 & \tau_1 \\
\tau_1 & |\tau |^2 \end{array}\right)
\end{equation}

 There is an action of
$SL(2,\R)$ under which an element $g\in SL(2,\R)$   given by
\begin{equation}
g=\left(\begin{array}{cc} a  & b
\\ c & d
\end{array}\right)    \qquad  ad-bc=1
\end{equation}
acts on the torus modulus  as
\begin{equation}
\tau\rightarrow\frac{a\tau+b}{c\tau+d} \equiv \gamma[\tau]
\end{equation}
In the $T^2$ bundle over $S^1$, the torus modulus $\tau$ varies with the circle coordinate $y$,
 with the $y$ dependence given by the $SL(2,\R)$ transformation
$\gamma(y)=exp\left(My\right)$, so that
\begin{equation}
\tau(y) =\gamma(y)[\tau_0]
\end{equation}
 for some fixed $\tau_0$.
 The metric on the bundle is then (\ref{metricbun}), which can be rewritten as  \cite{Hull:1998vy}
 \begin{equation}\label{torus metric}
ds^2=R^2dy^2+\frac{A}{Im(\tau)}|dz^1+\tau dz^2|^2
\end{equation}
with $\tau (y)=\gamma(y)[\tau_0]$ and constant $A$.

\subsubsection{Conjugacy Classes of $SL(2,\R)$ and $SL(2,\Z)$}

For the reduced theory truncated to the  sector independent of the internal coordinates, the monodromy
can be in $SL(2,\R)$, and the distinct theories are classified by $SL(2,\R)$ conjugacy classes   and there are three distinct theories corresponding to the three conjugacy classes of
$SL(2,\R)$ \cite{Hull:1998vy}.
If the massive Kaluza-Klein modes
are kept, then the monodromy must be in $SL(2,\Z)$, and there is a richer class of theories corresponding to the  conjugacy classes of $SL(2,\Z)$.

For $SL(2,\R)$ monodromy, the
three conjugacy classes are the elliptic, the parabolic and the
hyperbolic classes which have $|Tr(e^{M_e})|<2$, $|Tr(e^{M_p})|=2$
and $|Tr(e^{M_h})|>2$ respectively.
The three conjugacy classes of $SL(2,\R)$ are
\begin{equation}\label{UNU}
{\cal M}_e=Ue^{N_e}U^{-1} \qquad {\cal M}_h=Ue^{N_h}U^{-1} \qquad {\cal M}_p=Ue^{N_p}U^{-1}
\end{equation}
where the mass matrices are
\begin{equation}\label{mass matrices}
N_p=\left(\begin{array}{cc}0&m\\0&0\end{array}\right) \qquad
N_h=\left(\begin{array}{cc}m&0\\0&-m\end{array}\right) \qquad
N_e=\left(\begin{array}{cc}0&\theta\\-\theta&0\end{array}\right)
\end{equation}
where $\theta$ takes values in the range $[0,2\pi]$,  $m$ is real and $U$ is an
arbitrary matrix of $SL(2,\R)$.

For the monodromy ${\cal M}$ to be in $SL(2,\Z)$ requires \lq quantization conditions' on the parameters $m, \theta$ and restrictions on $U$.
For the parabolic class,
the restriction to $SL(2,\Z)$ requires $m\in\Z$
and the conjugacy classes are represented by
\begin{equation} {\cal
M}_m=\left(\begin{array}{cc}1&m\\0&1\end{array}\right)
\end{equation}
where $m\in\Z$ give a distinct conjugacy class for each integer
$m$.

For the elliptic class,
  if
$U=1$ then  $\theta=m\pi/2$ where $m$ is an integer, giving two classes represented by
\begin{equation}
{\cal M}_2=\left(\begin{array}{cc}-1&0 \\0&-1\end{array}\right)
\quad {\cal
M}_4=\left(\begin{array}{cc}0&1\\-1&0\end{array}\right)
\end{equation}
which generate the groups $\Z_2$ and $\Z_4$ respectively.
There are two more elliptic $SL(2,\Z)$
 conjugacy classes (with $U\ne 1$), represented by
\begin{equation}
{\cal M}_3=\left(\begin{array}{cc}0&1\\-1&-1\end{array}\right)
\quad{\cal M}_6=\left(\begin{array}{cc}1&1\\-1&0\end{array}\right)
\end{equation}
which generate the $\Z_3$ and $\Z_6$ groups respectively. These
are of the form
\begin{equation}
{\cal M}_3=UN_e(\theta=2\pi/3)U^{-1}    \qquad  {\cal
M}_6=UN_e(\theta=\pi/3)U^{-1}
\end{equation}
for certain $U$.

There is an infinite family of hyperbolic conjugacy classes with  $U\ne 1$
 given
by
\begin{equation}
{\cal M}_m=\left(\begin{array}{cc}m&1\\-1&0\end{array}\right)
\end{equation}
where $m\in\Z$ and $m>2$. In addition there are an infinite number
of sporadic monodromies ${\cal M}(t)$ where $t$ denotes the trace
of the matrix, again with $U\ne 1$ \cite{DeWolfe:1998eu}. The
first five are:
\begin{eqnarray}
{\cal M}(8)&=&\left(\begin{array}{cc}1&2\\3&7\end{array}\right)
\quad {\cal
M}(10)=\left(\begin{array}{cc}1&4\\2&9\end{array}\right) \quad
{\cal
M}(12)=\left(\begin{array}{cc}1&2\\5&11\end{array}\right)\nonumber\\
{\cal
M}(13)&=&\left(\begin{array}{cc}2&3\\7&11\end{array}\right)\quad
{\cal M}(14)=\left(\begin{array}{cc}1&1\\6&13\end{array}\right)...
\end{eqnarray}

\subsubsection{Parabolic Twist}

We shall consider the application of these monodromy matrices to
Scherk-Schwarz reductions. First, we consider a twist by an
element of the parabolic conjugacy class of $SL(2,\R)$ given by
\begin{equation}
{\cal M}_p=\left(\begin{array}{cc} 1  & m
\\ 0 & 1
\end{array}\right)
\end{equation}
for some fixed $m\in \R$. This will be in $SL(2,\Z)$ if the mass
parameter $m$ is quantized, $m \in \Z$. The Scherk-Schwarz ansatz
gives
\begin{equation}\label{Parabolic tau}
 \gamma_p(y)=\left(\begin{array}{cc}  1 & my
\\ 0 & 1
\end{array}\right)  \qquad \tau (y)=\tau_0+my
\end{equation}
where the mass matrix is
\begin{equation}
 M_p=\left(\begin{array}{cc}  0 & m
\\ 0 & 0
\end{array}\right)
\end{equation}
and $\tau_0=\tau_1+i\tau_2$ is some constant modulus. For simplicity we shall  choose $A=R=1$ in the metric (\ref{torus metric}).

If $m\in \Z$, this reduction may be thought of as a twisted torus
reduction on $\cG/\G$ where $\cG$ is the group manifold for the
Heisenberg group \cite{Kachru:2002sk}, as we shall now review.
This space is sometimes called the nilmanifold and the metric is
given by (\ref{torus metric}) with $\tau(y)$ given by
(\ref{Parabolic tau}). The generators satisfy the Heisenberg
algebra
\begin{equation}
[t_2,t_y]=mt_1 \qquad  [t_1,t_y]=0 \qquad  [t_1,t_2]=0
\end{equation}
and the group element
  $g\in {\cal G}$ corresponding to the coordinates $y, z^a$ is
\begin{equation}
g(y,z^a)= \left(\begin{array}{ccc} 1 & -my & z^1
\\ 0 & 1 & z^2 \\ 0 & 0 & 1
\end{array}\right)
\end{equation}
The Heisenberg group is non-compact and the
compact nilmanifold is obtained by identifying the coordinates
under
\begin{equation}
(y,z^1,z^2)\sim (y+\alpha,z^1-m\alpha z^2+\beta^1,z^2+\beta^2)
\end{equation}
with $\alpha,\beta^a \in \Z$. This can be understood as a
quotient by a discrete subgroup  $\Gamma\subset \cG_L$. This
identification may be written as the left quotient $g\sim h\cdot
g$ where $h\in \G$ and $\G$ is the discrete subgroup of matrices
of the form
\begin{equation}
h(\alpha,\beta^a)= \left(\begin{array}{ccc}  1 & -m\alpha &
\beta^1
\\ 0 & 1 & \beta^2 \\ 0 & 0 & 1
\end{array}\right)\in \Gamma
\end{equation}
with integer $\alpha,\beta^a$.  The left action $g\to
h\cdot g$ leaves the metric invariant, so the identification is
consistent with the ansatz.

\subsubsection{Elliptic Twist}

As a second example, we take $\mathcal{M}$ to lie in the elliptic
conjugacy class of $SL(2,\Z)$ where
\begin{equation}
\gamma_e(y)=U\left(\begin{array}{cc} \cos(\theta y) & \sin(\theta y) \\ -\sin(\theta y) & \cos(\theta y) \end{array}\right)  U^{-1} \qquad M_e=U \begin{pmatrix}0 & \theta  \\
                      -\theta & 0 \end{pmatrix}  U^{-1}
 \end{equation}
then the group ${\cal G}$ generated by $\{t_a,t_y\}$ is
$ISO(2)$, the group of isometries of the Euclidean plane and
$\theta$ takes values between $0$ and $2\pi$. The monodromy is
\begin{equation}
{\cal M}_e=U\left(\begin{array}{cc}\cos(\theta) & \sin(\theta) \\
-\sin(\theta) & \cos(\theta) \end{array}\right) U^{-1}
\end{equation}
and if $U=1$ the complex structure is
\begin{equation}
\qquad
\tau=\frac{\tau_0\cos(\theta y)+\sin(\theta y)}{-\tau_0\sin(\theta
y)+\cos(\theta y)}
\end{equation}
Note that if $\tau_0=i$, then $\tau=i$ is independent of $\theta$.

The metric of the internal space of the reduced theory is
$ds^2=(\sigma^y)^2+H(y)_{ab}\sigma^a\sigma^b$ where the
left-invariant one-forms are
$\sigma^m=(\sigma^y,\gamma_e(y)^a{}_bdz^b)$. The group manifold of
$ISO(2)$ has topology $S^1\times \R^2$ and is parameterised by
matrices of the form
\begin{equation}\label{elliptic g}
g(y,z^a)= {\cal U} \left(\begin{array}{ccc} \cos(\theta y) & -\sin(\theta
y) & \tilde z^1
\\ \sin(\theta y) & \cos(\theta y) & \tilde z^2 \\ 0 & 0 & 1
\end{array}\right) {\cal U}^{-1}
\end{equation}
where
\begin{equation}
 {\cal U}=\left(\begin{array}{cc} U &0 \\ 0& 1 \end{array}\right)
\end{equation}
and \begin{equation}
\tilde z^a=( U^{-1})^a{}_b z^b
\end{equation}

The space is compactified under the left action of
\begin{equation}
h(\alpha,\beta^a)=\left(\begin{array}{ccc} 1 & 0 & \beta^1 \\
0 & 1 & \beta^2 \\ 0 & 0 & 1
\end{array} \right)
\end{equation}
where we require $\beta^a\in\Z $.

The Cartan-Killing metric is degenerate for this case;
$\eta=diag\{1,0,0\}$ but there is an invariant metric of the form
$diag\{a,b,b\}$ for any $a,b$. For any choice of metric $g_{mn}$
on $\cX= T^3$, the vector fields $Z_1,Z_2$ generating the
non-compact part of $\cG$ will not be Killing vectors and the
gauge group $ISO(2)$ will be broken to at most the compact
subgroup $SO(2)$. Indeed, for this reduction the scalar potential
has a minimum  at $g_{mn}=\d _{mn}$ \cite{Dabholkar:2002sy} and in
this vacuum the gauge group $ISO(2)$ is spontaneously broken to
the $SO(2)$ generated by $Z_y$.

As a final comment, we note that if  $\tau_0= i$ the
$y$-dependence of $\tau$ cancels out in the above ansatz and
$\tau=\tau_0= i$ as $\tau_0= i$ is a fixed point of the $SL(2,\R)$
transformation generated by $M$. This   fixed point of the
T-duality twist   is a minimum of the Scherk-Schwarz potential
$V(\tau, A)$ at $\tau=i$, and   the theory at the minimum
corresponds to to an orbifold reduction for which there is an
exact conformal field theory description \cite{Dabholkar:2002sy}.

\subsubsection{Hyperbolic Twist}

The third and final conjugacy class  is the
hyperbolic with mass matrix $M$ and monodromy ${\cal M}$ given by
\begin{equation}
M_h=U\left(\begin{array}{cc} m  & 0
\\ 0 & -m
\end{array}\right) U^{-1} \qquad  {\cal M}_h=U\left(\begin{array}{cc} e^m  & 0
\\ 0 & e^{-m}
\end{array}\right) U^{-1}
\end{equation}
for some $U$.
The Scherk-Schwarz ansatz gives
dependence on the circle coordinate $y$ through
\begin{equation}
\gamma_h(y)=U\left(\begin{array}{cc} e^{my} & 0
\\ 0 & e^{-my}\end{array}\right) U^{-1}
\end{equation}
The Lie algebra is that of $ISO(1,1)$
 and the
group elements are
\begin{equation}\label{hyperbolic g}
g(y,z^a)= {\cal U} \left(\begin{array}{ccc} e^{-my} & 0 & \tilde z^1
\\ 0 & e^{my} & \tilde z^2 \\ 0 & 0 & 1
\end{array}\right) {\cal U}^{-1}
\end{equation}
and a compact space is obtained by identifying under the left
action of a discrete subgroup $\Gamma\subset{\cal G}$ of
matrices of the form
\begin{equation}
h(\alpha,\beta^a)=\left(\begin{array}{cc} {\cal M}^{-\alpha} &
\beta \\ 0 & 1 \end{array}\right)
\end{equation}
where   $\alpha,\beta^a$ are integers. The
left-invariant Maurer-Cartan forms and generators of the right
action are well defined on the compact space.

%%%%%%%%%%%

\section{Scherk-Schwarz Dimensional Reduction}

In this section, we review the results of Scherk-Schwarz
 dimensional reduction~\cite{Scherk:1979zr} in field theory, truncating to the zero-mode sector, and include fluxes,
 generalising the results of \cite{Odd}. It can be viewed as a compactification on
  a $d$-dimensional compact
space ${\cal X}={\cal G}/\Gamma$ given by the left-coset of a
group manifold $\cG$ by a discrete subgroup $\G$, as in the last
section, and
 much of the structure of the theory
reduced on ${\cal X}$ is the same as it would be for the reduction
on the group manifold ${\cal G}$,  and in particular there is a
consistent truncation to a finite set of fields in the reduced
theory. The ansatz uses the one-forms $\sigma^m= \sigma _i{}^m
dy^i$ satisfying  the structure equation
\begin{equation}
 d\sigma^m+\frac{1}{2}f^m_{np}\sigma^n
\wedge \sigma^p=0
\end{equation}

\subsection{Gravity Reduction}

We shall reduce the $D$-dimensional Einstein-Hilbert Lagrangian
\begin{equation}
{\cal L}_D=\widehat{R}*1
\end{equation}
on a $d$-dimensional   manifold ${\cal X}={\cal G}/\Gamma$,
following  the notation of \cite{Cvetic:2003jy}. The spacetime has
coordinates $\{x^{\mu},y^i\}$, where the $y^i$ are coordinates for
the internal space $\mathcal{X}$ and the $x^\mu$ are the
coordinates for the reduced spacetime. The most general
left-invariant Einstein frame reduction ansatz is\footnote{Note
$\varphi$ introduced in the ansatz is distinct from the dilaton
$\phi$ that will be introduced when we come to consider the string
frame Lagrangian.}
\begin{equation}
ds^2_D=e^{2\alpha\varphi}ds^2_d+e^{2\beta\varphi}g_{mn}\nu^m\nu^n
\end{equation}
where the one-forms
\begin{equation}
\nu^m=\sigma^m - A^m
\end{equation}
introduce the Kaluza-Klein gauge fields $A^m_\m$, which have
two-form field strength
\begin{equation}
F^m=dA^m+\frac{1}{2}f^m_{np}A^n\wedge A^p
\end{equation}
We have retained the $d$ ${\cal G}_L$ invariant Kaluza-Klein gauge
fields (graviphotons) $A^m(x)=A^m_{\mu}(x)dx^{\mu}$ and the
$d(d+1)/2$ scalars $g_{mn}(x)$ and $\varphi$, where $g_{mn}(x)$
has unit determinant. All other fields in the gravity sector are
truncated out in the Scherk-Schwarz ansatz. Note that the ansatz
breaks the ${\cal G} _L\times {\cal G}_R$ symmetry of ${\cal G}$
down to ${\cal G} _L$ unless $g_{mn}(x)$ is an invariant metric,
which for semi-simple groups requires it to be proportional to
$\eta_{mn}$.

The condition for the volume element ($\sqrt g$) to be invariant under the left action $\cG_L$
is
\begin{equation}
f^m_{nm}=0
\end{equation}
which is the condition that the group be unimodular (i.e. the
adjoint action on the Lie algebra is trace-free). If this is
satisfied, then the action can be dimensionally reduced to give an
action that is independent of the internal coordinates
\cite{Scherk:1979zr}. If this condition is not satisfied, one can
instead dimensionally reduce the field equations, giving a set of
field equations that are independent of the internal coordinates,
but which in general cannot be derived from an action that is
independent of the internal coordinates \cite{Bergshoeff:2003ri}.
Thus for unimodular groups, there is a consistent truncation of
the action, while for groups which are not unimodular, there is  a
consistent truncation of the field equations, but in general not
of the action. If one keeps the full Kaluza-Klein theory without
truncation, there is no need to apply this condition and, for
compact  internal space, there will still be a mass gap, and there
will an infinite tower of massive field equations on the reduced
space. Here we will present results for the reduction of the
action for unimodular $\cG$. There is a generalisation of our
results to the reduction of the field equations in the
non-unimodular case (see. \cite{Bergshoeff:2003ri} for examples).
For reductions with duality twists for which $\cG$ has Lie algebra
(\ref{duality twist algebra}), the group will be unimodular if the
mass matrix is traceless, $M_i{}^i=0$.

The reduced Lagrangian is \cite{Scherk:1979zr,Cvetic:2003jy,Odd}
\begin{eqnarray}
{\cal L}_D&=&R*1 - \frac{1}{2}*d\varphi\wedge d\varphi -
\frac{1}{2}g^{mp}g^{nq}*Dg_{mn}\wedge Dg_{pq}
-\frac{1}{2}e^{2(\beta-\alpha)\varphi}g_{mn}*F^m \wedge F^n
\nonumber\\
&&-\frac{1}{2}e^{2(\beta-\alpha)\varphi}\left(
g_{mn}g^{pq}g^{ts}f^m_{pt}f^n_{qs}+2g^{mn}f^p_{qm}f^q_{pn}\right)*1
\end{eqnarray}
where
\begin{equation}\alpha=-\left( \frac{D-d}{2(d-2)(D-2)}\right)^{\frac{1}{2}}
\qquad \beta=\left( \frac{d-2}{2(D-d)(D-2)}\right)^{\frac{1}{2}}
\end{equation}
and
\begin{equation}
Dg_{mn}=dg_{mn}+g_{mp}f^p_{nq}A^q+g_{np}f^p_{mq}A^q
\end{equation}

\subsection{Antisymmetric Tensor Gauge Field Reduction with Flux}

In this section we consider the reduction of a p-form gauge field
$\widehat{B}_{(p)}$ with $p+1$-form field strength
$\widehat{G}_{(p+1)}=d\widehat{B}_{(p)}$. We include the most
general $\cG_L$-invariant flux for the   field strength
$\widehat{G}_{(p+1)}$,
\begin{equation}
\widehat{G}_{(p+1)}=\frac{1}{(p+1)!}K_{mn...p+1} \s^m\wedge
\s^n\wedge ....\wedge\s^{p+1}+\dots
\end{equation}
where $K_{m_1 m_1...m_{p+1}}$ are constant coefficients. The
Bianchi identity $d\widehat{G}_{(p+1)}=0$, and therefore
$d(K_{mn...p}\s^m\wedge \s^n\wedge ....\wedge\s^{p+1})=0$, imposes
the integrability condition
\begin{equation}
K_{[m_1 m_2 m_3 ...m_p|n}f^n_{|m_{p+1} m_{p+2}]}=0
\end{equation}
and we require the constant $K_{mn...p+1}$ to satisfy this
algebraic constraint. We use the most general ${\cal G} _L$
-invariant reduction ansatz for $\widehat{B}_{(p)}$, which  is
\begin{eqnarray}
\widehat{B}_{(p)}&=&B_{(p)}+B_{(p-1)m}\wedge\nu^m+\frac{1}{2!}B_{(p-2)mn}\wedge
\nu^m\wedge\nu^n+...\nonumber\\
&&...+\frac{1}{p!}B_{(0)m_1 m_2
m_3...m_p}\nu^{m_1}\wedge...\wedge\nu^{m_p}+\varpi_{(p)}
\end{eqnarray}
where we have included  the   flux $\widehat{G}_{m_1 m_1...m_p+1}=
(-)^pK_{m_1 m_1...m_{p+1}}$ through $\varpi_{(p)}$, which
satisfies
\begin{equation}
d\varpi_{(p)}=\frac{1}{(p+1)!}K_{m_1
m_2...m_{p+1}}\sigma^{m_1}\wedge...\wedge\sigma^{m_{p+1}}
\end{equation}
We use this notation, even though the flux $\varpi_{(p)}$ may not
be defined globally, to emphasise the requirement
$d(K_{mn...p+1}\s^m\wedge \s^n\wedge ....\wedge\s^{p+1})=0$. Note
that the spectrum is the same as for a toroidal reduction, with
one $p$-form, $d$ {} $p-1$-forms, $d(d-1)/2$ {} $p-2$-forms etc,
but these are now charged under the gauge group of the reduced
theory in general.

The field strength $\widehat{G}_{(p+1)}$ can then be decomposed as
\begin{eqnarray}
\widehat{G}_{(p+1)}&=&G_{(p+1)}+G_{(p)m}\wedge\nu^m+\frac{1}{2!}G_{(p-1)mn}\wedge
\nu^m\wedge\nu^n+...\nonumber\\
&&...+\frac{1}{(p+1)!}G_{(0)m_1 m_2
m_3...m_{p+1}}\nu^{m_1}\wedge...\wedge\nu^{m_{p+1}}
\end{eqnarray}
where, the $i$-form field strengths are
\begin{eqnarray}
G_{(i)m_1 m_2...m_{p+1-i}}&=&DB_{(i-1)m_1 m_2...m_{p+1-i}}+(-)^p
B_{(i-2)m_1 m_2...m_{p+1-i}n}\wedge F^n
\nonumber\\
&&+(-)^pc(i,p)f^n_{[m_1 m_2}B_{(i)m_3 m_4...m_{p+1-i]}n} \nonumber\\
&&+\frac{(-)^p}{i!}K_{m_1 m_2...m_{p+1}}A^{m_{p+1-i}}\wedge
A^{m_{p+2-i}}\wedge...\wedge A^{m_{p+1}}
\end{eqnarray}
and the coefficient $c(i,p)$ is given by
\begin{equation}
c(i,p)=\frac{(p+1-i)!}{2(p-1-i)!}
\end{equation}
The ${\cal G}_L$-covariant derivatives are
\begin{equation}
DB_{(i-1)m_1...m_{p+1-i}}=dB_{(i-1)m_1...m_{p+1-i}}+(-)^ic(i,p)B_{(i-1)[m_1
m_2...m_{p-i}|n}f^n_{|m_{p-i+1}]q}\wedge A^q
\end{equation}
The generic antisymmetric tensor Lagrangian
\begin{equation}
\mathcal{L}_{\widehat{G}}=-\frac{1}{2}*\widehat{G}_{(p+1)}\wedge\widehat{G}_{(p+1)}
\end{equation}
is therefore reduced to
\begin{equation}
\mathcal{L}_{\widehat{G}}=-\frac{1}{2}\sum_{i=0}^{p+1}g^{m_1m_2}g^{m_3m_4}...g^{m_{2(p+1-i)-1}m_{2(p+1-i)}}
*G_{(i)m_1m_3...m_{2(p+1-i)-1}}\wedge G_{(i)m_2m_4...m_{2(p+1-i)}}
\end{equation}

%%%%%%%%%

\section{String Theory Compactifications on Twisted Tori}

If a group $\cG$ has a cocompact discrete subgroup $\G$, then
we have seen that the Scherk-Schwarz reduction of a field theory
using the structure constants of $\cG$
is equivalent to the compactification on  a  compact twisted torus ${\cal
G}/\Gamma$ (or on the group manifold $\cG$ itself, if this is compact) with $\cG_L$-invariant ansatz for the ground state fields,
followed by a consistent truncation to a $\cG_L$-invariant sector of the spectrum that is independent of the internal coordinates. The resulting theory has a
local gauge symmetry that includes the isometry group $\cG_R$.
As the internal space is compact, this can be extended to the full Kaluza-Klein theory
by compactifying on ${\cal
G}/\Gamma$ with  $\cG_L$-invariant metric and fluxes, but keeping the full Kaluza-Klein spectrum, which will have a tower of massive states separated by a mass-gap.

This can now be extended to string theory or M-theory by
considering compactification   on the compact space ${\cal
G}/\Gamma$ with a $\cG_L$-invariant ansatz for the metric and anti-symmetric tensor gauge fields.
Again there is a well-defined mass-gap, and the theory
can be truncated to the Kaluza-Klein theory, which in turn can be truncated to the
$\cG_L$-invariant sector and to the fields independent of the internal coordinates,   corresponding to the Scherk-Schwarz reduction of the low-energy field theory.
This effective  field theory is a consistent truncation and for generic vacua includes all the light fields.
However, for special vacua, the symmetry is enhanced and there are extra light fields.
For example, for generic vacua, the metric gives gauge fields for the gauge group $\cG_R$, but if the
metric is chosen to be a $\cG_L\times\cG_R$-invariant metric (the Cartan-Killing metric for a compact group) then the metric gives massless gauge fields for the gauge group $\cG_L\times\cG_R$.
In the following subsections, we will describe the Scherk-Schwarz reduction of the low-energy field theory.

Thus a Scherk-Schwarz reduction of a low-energy field theory
associated with a group $\cG$ can be extended to a Kaluza-Klein
compactification or a compactification of string theory if there
is a  discrete subgroup $\G\subset \cG_L$ such that $\cG/\G$ is
compact, i.e. if there is a cocompact discrete subgroup. A
necessary condition for this is that the group be unimodular, i.e.
that the structure constants satisfy $f_{nm}^m=0$. Note that
semi-simple groups are unimodular, and the groups with algebra
 (\ref{duality twist algebra}) arising in reductions with duality twists are unimodular if the mass matrix is traceless, $M_i{}^i=0$.
For non-unimodular  groups $\cG$,  as discussed in section 4, although there is no Scherk-Schwarz reduction
of the action,
there is a Scherk-Schwarz reduction of the field equations, but  this cannot
be extended to a compactification of string theory.

%%%%%%%%%%%

\subsection{Scherk-Schwarz Reduction of Low Energy Field Theory}

We shall apply the results of the last section to the reduction of
the $D$-dimensional Lagrangian
\begin{equation}
\label{string frame lagrangian} {\cal
L}_S=e^{-\widehat{\Phi}}\left( \widehat{R}*1+*d\widehat{\Phi}
\wedge d\widehat{\Phi} - \frac{1}{2}*\widehat{G}_{(3)}\wedge
\widehat{G}_{(3)} \right)
\end{equation}
governing  the massless fields of the bosonic string, or a subset
of the massless  bosonic fields of the various superstrings. It is
related to the Einstein frame Lagrangian by a conformal scaling
\begin{equation}
ds^2_{String}=e^{-a\phi}ds^2_{Einstein}
\end{equation}
where $\widehat{\Phi}=-\frac{1}{2a}\phi$, $a^2=8/(D-2)$. The
Lagrangian in the Einstein frame is
\begin{equation}
{\cal L}_E=\widehat{R}*1 - \frac{1}{2}*d\phi \wedge d\phi -
\frac{1}{2}e^{a\widehat{\phi}}*\widehat{G}\wedge \widehat{G}
\end{equation}
where $\widehat{G}=d\widehat{B}$ and $\widehat{B}$ is the
Kalb-Ramond form.

The Scherk-Schwarz reduction of this theory was given in
\cite{Scherk:1979zr}, and the generalization to include flux was
considered in \cite{Cvetic:2003jy,Odd}. We introduce a  flux
$\widehat{G}=K+\dots$ where
\begin{equation}
K=\frac{1}{6}K_{mnp}\sigma^m\wedge\sigma^n\wedge\sigma^p
\end{equation}
with constant coefficients $K_{mnp}$, and requiring $dK=0$
 gives the integrability
condition
\begin{equation}
\label{eq:B-field integrability}K_{t[mn }f^t_{pq]}=0
\end{equation}
which will later arise as part of the  Jacobi identities for the
gauge algebra of this theory. As $K$ is closed, there is locally a
2-form potential $\varpi_{(2)}$ for the flux such that
\begin{equation} d\varpi_{(2)}=K
\end{equation}
so that the ansatz for the potential is of the form
$\widehat{B}=\varpi_{(2)}+\dots$. The ansatz  for the
reduction of the  potential is
\begin{equation}\label{ansatz}
\widehat{B}=B_{(2)}+B_{(1)m}\wedge\nu^m+\frac{1}{2}B_{(0)mn}\nu^m\wedge\nu^n+\varpi_{(2)}
\end{equation}
giving the field strength
\begin{equation}
\widehat{G}=G_{(3)}+G_{(2)m}\wedge\nu^m+\frac{1}{2}G_{(1)mn}\wedge\nu^m\wedge\nu^n+\frac{1}{6}G_{(0)mnp}\nu^m\wedge\nu^n\wedge\nu^p
\end{equation}
where
\begin{eqnarray}\label{G field strengths}
G_{(3)}&=&dB_{(2)}+B_{(1)m}\wedge F^m+\frac{1}{6}K_{mnp}A^m\wedge
A^n\wedge A^p
\nonumber\\
G_{(2)m}&=&DB_{(1)m}+B_{(0)mn}F^n+\frac{1}{2}K_{mnp}A^n\wedge A^p
\nonumber\\
G_{(1)mn}&=&DB_{(0)mn}+f^p_{mn}B_{(1)p}+K_{mnp}A^p
\nonumber\\
G_{(0)mnp}&=&3B_{(0)[m|q}f^q_{|np]}+K_{mnp}
\end{eqnarray}
and
\begin{eqnarray}
DB_{(1)m}&=&dB_{(1)m} - B_{(1)n}f^n_{mp}\wedge A^p
\nonumber\\
DB_{(0)mn}&=&dB_{(0)mn}+2B_{(0)[m|p}f^p_{|n]q}A^q
\end{eqnarray}
The Lagrangian of the Kalb-Ramond sector of the reduced theory, in
the Einstein frame, is then
\begin{eqnarray}
{\cal L}_D&=&e^{a\phi-4\alpha\varphi}\left(-\frac{1}{2}*G_{(3)}
\wedge G_{(3)} -
\frac{1}{2}e^{-2(\beta-\alpha)\varphi}g^{mn}*G_{(2)m}\wedge
G_{(2)n} \right.
\nonumber\\
&&\left.-\frac{1}{2}e^{-4(\beta-\alpha)\varphi}g^{mn}g^{pq}*G_{(1)mp}\wedge
G_{(1)nq}   \right.
\nonumber\\
&&\left.-
\frac{1}{2}e^{-6(\beta-\alpha)\varphi}g^{mn}g^{pq}g^{ts}*G_{(0)mpt}\wedge
G_{(0)nqs}\right)
\end{eqnarray}
It will be convenient to work in the string frame, where the
metric reduction ansatz is
\begin{equation}
\label{straing frame metric}
ds^2_D=g_{\mu\nu}dx^{\mu}dx^{\nu}+g_{mn}\nu^m\nu^n
\end{equation}
Reducing (\ref{string frame lagrangian}) on the spacetime of
(\ref{straing frame metric}) gives
\begin{eqnarray}
\label{B-field lagrangian} {\cal
L}_D&=&e^{-\phi}\left(R*1+*d\phi\wedge
d\phi+\frac{1}{2}*Dg^{mn}\wedge
Dg_{mn}-\frac{1}{2}g_{mn}*F^m\wedge F^n -\frac{1}{2}*G_{(3)}\wedge
G_{(3)} \right.
\nonumber\\
&&\left.- \frac{1}{2}g^{mn}*G_{(2)m}\wedge
G_{(2)n}-\frac{1}{2}g^{mn}g^{pq}*G_{(1)mp}\wedge
G_{(1)nq}+V*1\right)
\end{eqnarray}
where the potential $V$ is
\begin{equation}
V=- \frac{1}{4}g_{mn}g^{pq}g^{st}f^m_{ps}f^n_{qt} -
\frac{1}{2}g^{mn}f^p_{mq}f^q_{np} -
\frac{1}{2}g^{mn}g^{pq}g^{st}G_{(0)mps}G_{(0)pqt}
\end{equation}
Here $\phi$ is  the shifted dilaton
\begin{equation}
\phi=\widehat{\Phi}-ln(\sqrt{g})
\end{equation}
where $g=det(g_{mn})$.

\subsection{Gauge algebra}

In this section we consider the gauge algebra of the reduced
Lagrangian (\ref{B-field lagrangian}). The gauge fields of the
reduced theory are the 2-form $B_{(2)}$, the $d$ vector fields
$B_{(1)m}$ and the
 $d$ vector fields $A^m$
 and we denote the generators of the corresponding gauge transformations $W,X,Z$ and the parameters
  $\Lambda_{(1)}=\Lambda_{(1)}(x^{\mu})$, $\lambda_{(0)m}=\lambda_{(0)m}(x^{\mu})$ and $\omega^m=\omega^m(x^{\mu})$ respectively, so that the gauge transformations include
  \begin{eqnarray}
\delta_{W}(\Lambda_{(1)})B_{(2)}&=&d\Lambda_{(1)}
\nonumber\\
\delta_{X}(\lambda_{(0)m})B_{(1)m}&=&d\lambda_{(0)m}+\dots
\nonumber\\
\delta_{Z}(\omega)A^m&=&-d\omega^m+\dots
\end{eqnarray}

The $W$ and $X$ symmetries come from
 the $D$-dimensional gauge transformations
$\widehat{B}\rightarrow \widehat{B}+d\widehat{\lambda}$ on
reducing the parameter
\begin{eqnarray}
\widehat{\lambda}&=&\Lambda_{(1)}+\lambda_{(0)m}\nu^m
\nonumber\\
d\widehat{\lambda}&=&d\Lambda_{(1)} - \lambda_{(0)m} F^m +
\left(d\lambda_{(0)m} + \lambda_{(0)n}f^n_{mp}A^p \right) \wedge
\nu^m - \frac{1}{2}\lambda_{(0)p}f^p_{mn}\nu^m \wedge\nu^n
\nonumber\\
\end{eqnarray}
The resulting  gauge transformations of the reduced potentials are
\begin{eqnarray}
\label{eq:B-field gauge}
\delta_{W}(\Lambda_{(1)})B_{(2)}&=&d\Lambda_{(1)}
\nonumber\\
\delta_{X}(\lambda_{(0)m})B_{(2)}&=&-\lambda_{(0)m}F^m
\nonumber\\
\delta_{X}(\lambda_{(0)m})B_{(1)m}&=&d\lambda_{(0)m}+\lambda_{(0)n}f^n_{mp}A^p=D\lambda_{(0)m}
\nonumber\\
\delta_{X}(\lambda_{(0)m})B_{(0)mn}&=&-\lambda_{(0)p}f^p_{mn}
\end{eqnarray}

The $Z$ symmetries arise from the diffeomorphism symmetry of the
higher dimensional theory. Under a diffeomorphism  of the internal
space with parameter $\omega^m(x)$
 the basis forms
transform as $\delta(\omega)\nu^m={\cal
L}_{\omega}\nu^m=\left(\iota_{\omega}d+d\iota_{\omega}\right)\nu^m=-\nu^nf^m_{np}\omega^p$.
The requirement $\delta(\omega)\widehat{B}=0$ that the ansatz
(\ref{ansatz}) is invariant under these general coordinate
transformations
 induces the following
transformations on the reduced potentials
\begin{eqnarray}\label{B-field diffeo}
\delta (\omega)B_{(2)}&=&\frac{1}{2}K_{mnp}\omega^pA^m\wedge
A^n+\left( d\Xi_{(1)}-\Xi_{(0)m}F^m \right)
\nonumber\\
\delta (\omega)B_{(1)m}&=&B_{(1)n}f^n_{mp}\omega^p -
K_{mnp}\omega^pA^n+D\Xi_{(0)m}
\nonumber\\
\delta
(\omega)B_{(0)mn}&=&2B_{(0)[m|p}f^p_{|n]q}\omega^q+K_{mnp}\omega^p-\Xi_{(0)p}f^p_{mn}
\end{eqnarray}
where
$\widehat{\Xi}=\Xi_{(1)}+\Xi_{(0)m}\nu^m=\iota_{\omega}\varpi_{(2)}$
has explicit internal coordinate dependence. We remove the
internal dependence by a gauge transformation
$\widehat{B}\rightarrow \widehat{B}+d\widehat{\lambda}$ with
parameter $\widehat{\lambda}=-\widehat{\Xi}$, yielding the gauge
transformations
\begin{eqnarray}
\label{mod diffeo}
\delta_{Z}(\omega)B_{(2)}&=&\frac{1}{2}K_{mnp}\omega^pA^m\wedge
A^n
\nonumber\\
\delta_{Z}(\omega)B_{(1)m}&=&B_{(1)n}f^n_{mp}\omega^p -
K_{mnp}\omega^pA^n
\nonumber\\
\delta_{Z}(\omega)B_{(0)mn}&=&2B_{(0)[m|p}f^p_{|n]q}\omega^q+K_{mnp}\omega^p
\nonumber\\
\delta_{Z}(\omega)A^m&=&-d\omega^m -
f^m_{np}\omega^pA^n=-D\omega^m \nonumber\\
\delta_{Z}(\omega)g_{mn}&=&2g_{(m|p}f^p_{|n)q}\omega^q
\end{eqnarray}
%We shall use this modified form of the diffeomorphisms to
%calculate the algebra and shall, in future, refer to the
%transformations (\ref{mod diffeo}) simply as the diffeomorphisms.

It is straightforward to calculate the gauge algebra of the
reduced theory using   the Jacobi identity $f^q_{[mn}f^t_{p]q}=0$
and the integrability condition $K_{t[mn }f^t_{pq]}=0$, giving
\begin{eqnarray}
\label{B-field algebra} \lbrack
\delta_{Z}(\widetilde{\omega}),\delta_{Z}(\omega)
\rbrack&=&\delta_{Z}(f^m_{np}\omega^n\widetilde{\omega}^p) -
\delta_{X}(K_{mnp}\omega^n\widetilde{\omega}^p) -
\delta_{W}(K_{mnp}\omega^n\widetilde{\omega}^pA^m)
\nonumber\\
\lbrack \delta_{X}(\lambda),\delta_{Z}(\omega)
\rbrack&=&-\delta_{X}(\lambda_mf^m_{np}\omega^p)
\end{eqnarray}
All other commutators vanish. Note the appearance of a
field-dependent parameter
$\Lambda_{(1)}=K_{mnp}\omega^n\widetilde{\omega}^pA^m$ for the
gauge transformation on the right hand side in (\ref{B-field
algebra}). This extra term involving an anti-symmetric tensor
gauge transformation only occurs when acting on the anti-symmetric
tensor potential $B_{(2)}$, and is absent when acting on the
anti-symmetric tensor field strength $G_{(3)}$. Such a
field-dependent term in the algebra is a natural feature of
theories with Chern-Simons terms, as we
 shall discuss   in the
next section.

The Lie algebra underlying this field-dependent gauge algebra is
the algebra found in~\cite{Odd}
\begin{eqnarray}
\label{B-field eqiv. class algebra}
\left[Z_m,Z_n\right]&=&f^p_{mn}Z_p-K_{mnp}X^p
\nonumber\\
\left[X^m,Z_n\right]&=&-f^m_{np}X^p
\nonumber\\
\left[X^m,X^n\right]&=&0
\end{eqnarray}
where, following  ~\cite{Odd}, we have defined the generators
$Z_m,X^m$ of the spin-one transformations
$\delta_{Z}(\omega)=\omega^mZ_m$ and
$\delta_{X}(\lambda_m)=\lambda_mX^m$. The Jacobi identity of this
algebra is equivalent to the integrability  conditions
$f^q_{[mn}f^t_{p]q}=0$ and $K_{t[mn }f^t_{pq]}=0
$. Note that in
the absence of flux, this contains  the algebra of the group $
{\cal G}_d$ generated by the
  right action vector fields $Z_m$
of section 2 and the
 $2d$ dimensional gauge group is  ${\cal
G}_{2d}\simeq {\cal G}_d\ltimes U(1)^d$. However,  the presence of
flux modifies the commutator of the $Z_m$ and can lead to
non-trivial algebras, although, as we shall see in section 5, in
some cases the algebra simplifies.

\subsection{$O(d,d)$ Covariant Formulation}

Remarkably, the theory can be written in an $O(d,d)$ covariant
form~\cite{Odd,Maharana:my,Giveon:1988tt}. The Lie algebra
(\ref{B-field eqiv. class algebra}) may be written as
\begin{equation}\label{Odd lie algebra}
\left[ T_A,T_B \right]=t_{AB}{}^CT_C
\end{equation}
where the   generators $Z_m, X^m$ ($m=1,2,3...d$) are combined
into an $O(d,d)$ vector
\begin{equation}
T_A= \left(\begin{array}{cc} Z_m & X^m
\end{array}\right)
\end{equation}
with  $A=1,2,3...2d$. Defining $t_{ABC}=L_{AD}t_{BC}{}^D$ where
$L_{AB}$
  is the  $O(d,d)$ invariant matrix
  \begin{equation} L_{AB}=\left(\begin{array}{cc}
0 & \bid_d \\ \bid_d & 0
\end{array}\right) \end{equation}
 the structure constants are $t
_{np}{}^m=f^m_{np}$ and $t_{mnp}=K_{mnp}$. $\bid_d$ is the
d-dimensional identity matrix $\delta_{mn}$.

To write the full Lagrangian in a manifestly $O(d,d)$covariant
form we define ~\cite{Odd,Maharana:my}
\begin{eqnarray}
{\cal B}_{(2)}&=&B_{(2)} - \frac{1}{2}B_{(1)m}\wedge A^m
\end{eqnarray}
and
\begin{equation}  {\cal A}^A= \left(\begin{array}{cc}
A^m \\ B_{(1)m}
\end{array}\right)  \qquad  {\cal F}^A= \left(\begin{array}{cc}
F^m \\ G_{(2)m} - B_{(0)mn}F^n
\end{array}\right)
\end{equation}
The scalars take values in the coset $  O(d,d)/O(d)\times
O(d)$
and can be parameterised by a coset metric
\begin{equation} {\cal M}^{AB}\equiv\left(\begin{array}{cc}
{\cal M}^{mn} & {\cal M}_n{}^m \\ {\cal M}_m{}^n & {\cal M}_{mn}
\end{array}\right)= \left(\begin{array}{cc}
g^{mn} & -B_{(0)np}g^{pm} \\ -B_{(0)mp}g^{np} &
g_{mn}+g^{pq}B_{(0)mp}B_{(0)nq}
\end{array}\right)
\end{equation}
The reduced Lagrangian is then
\begin{eqnarray}\label{Odd Lagrangian} {\cal
L}_D&=&e^{-\phi}\left(R*1+*d\phi\wedge
d\phi+\frac{1}{2}*G_{(3)}\wedge
G_{(3)}+\frac{1}{4}L_{AC}L_{BD}*D{\cal M}^{AB}\wedge D{\cal
M}^{CD}\right.
\nonumber\\
&&- \left.\frac{1}{2}L_{AC}L_{BD}{\cal M}^{AB}*{\cal
F}^C\wedge{\cal F}^D- \frac{1}{12}{\cal M}^{AD}{\cal M}^{BE}{\cal
M}^{CF}t_{ABC}t_{DEF}\right. \nonumber\\
&&\left.+ \frac{1}{4}{\cal
M}^{AD}L^{BE}L^{CF}t_{ABC}t_{DEF}\right)
\end{eqnarray}
where
\begin{equation}\label{Chern-simons field strength}
G_{(3)}=d{\cal B}_{(2)}+\frac{1}{2}\left(L_{AB}{\cal A}^A\wedge
{\cal F}^B - \frac{1}{3!}t_{ABC}{\cal A}^A\wedge {\cal A}^B\wedge
{\cal A}^C\right)
\end{equation}
and
\begin{equation}
D{\cal M}^{AB}=d{\cal M}^{AB}+{\cal M}^{AC}t_{CD}{}^B{\cal
A}^D+{\cal M}^{BC}t_{CD}{}^A{\cal A}^D
\end{equation}
Note that
\begin{equation}
D{\cal M}_n{}^m= g^{mp}f^q_{np}B_{(1)q}+g^{mp}K_{npq}A^q+...
\end{equation}
and  the scalar kinetic term $(D{\cal M})^{2}$ term gives
  mass terms for the one-form fields $B_{(1)m}$ and $A^m$.

 This form of the theory is invariant under $O(d,d)$ transformations, provided the
 structure constants are taken to change under  $O(d,d)$, so that the
 $t_{AB}{}^C$ transform as a tensor under $O(d,d)$.

This Lagrangian is invariant under the gauge symmetry generated by
$\delta_Z(\omega)$, $\delta_X(\lambda)$ and $\delta_W(\Lambda)$.
Combining the gauge parameters $\omega^m, \lambda_m$ into an
$O(d,d)$ vector $\alpha^A=(\omega^m, -\lambda_m)^T$, the
transformation of the one-form gauge fields is
\begin{eqnarray}
\delta(\alpha){\cal A}^A=\alpha^BT_B:{\cal
A}^A&=&-d\alpha^A-t_{BC}{}^A\alpha^C{\cal
A}^B\nonumber\\
&=&\left(\begin{array}{cc} -d\omega^m-f^m_{np}\omega^pA^n \\
d\lambda_m+\lambda_nf^n_{mp}A^p+B_{(1)n}f^n_{mp}\omega^pA^n-K_{mnp}\omega^pA^n
\end{array}\right)
\end{eqnarray}
The 3-form field strength
\begin{equation}\label{G_3}
G_{(3)}=d{\cal B}_{(2)}+\frac{1}{2}\Omega({\cal A},{\cal F})
\end{equation}
has a
  Chern-Simons
form
\begin{equation}\label{Chern-simons form}
\Omega({\cal A},{\cal F})=L_{AB}{\cal A}^A\wedge {\cal F}^B -
\frac{1}{3!}t_{ABC}{\cal A}^A\wedge {\cal A}^B\wedge {\cal A}^C
\end{equation}
 satisfying
\begin{equation}\label{domegaFF}
d\Omega=L_{AB}{\cal F}^A\wedge{\cal F}^B
\end{equation}
As usual for field strengths with Chern-Simons terms, for
$G_{(3)}$ to be gauge invariant under the infinitesimal gauge
transformations $\delta(\alpha)$, it is necessary that ${\cal
B}_{(2)}$ transforms non-trivially. The required ${\cal B}_{(2)}$
transformation is, up to a total derivative,
\begin{equation} \delta(\alpha){\cal
B}_{(2)}=\frac{1}{2}L_{AB}\alpha^A d{\cal A}^B
\end{equation}
 The algebra of these transformations is field-dependent:
 \begin{eqnarray}\label{OddAlgebra}
[\delta(\widetilde{\alpha}),\delta(\alpha)]=\delta(t_{BC}{}^A\alpha^B\widetilde{\alpha}^C)
-\delta_W( t_{[ABC]} \alpha^B\widetilde{\alpha}^C{\cal A}^A)
\end{eqnarray}
This is simply the $O(d,d)$ covariant form of the algebra
(\ref{B-field algebra}). All reductions with non-trivial fluxes
and twists may be written in this form so one expects such a field
dependent algebra in any such reduction where
$t_{[ABC]}\neq 0$. As
usual, we shall refer to the Lie algebra (\ref{Odd lie algebra})
as the gauge algebra, even though strictly speaking the symmetry
algebra is the algebra (\ref{OddAlgebra}) with field-dependent
structure functions.

In the abelian limit in which the structure constants $t_{AB}{}^C$
are set to zero, this reduces to the standard reduction on a
$d$-torus, giving a  field theory with manifest global $O(d,d)$
symmetry. The non-abelian reduction discussed here can be thought
of as a gauging of a $2d$-dimensional subgroup ${\cal G}_{2d}$ of
the $O(d,d)$ symmetry.
In particular, when a supergravity theory is dimensionally reduced in this way, the result is a gauged supergravity theory with gauge group ${\cal G}_{2d}$.

%%%%%%%%%%%%%%%%%%%%

\section{Symmetry Breaking and Examples of Flux Reductions}

\subsection{Symmetry Breaking}

The twisted reduction with flux gives rise to a compactified
theory with $2d$-dimensional gauge group $\cG _{2d}$ with gauge
algebra (\ref{B-field eqiv. class algebra}), and this symmetry is
in general spontaneously broken. First, some of the gauge symmetry
is non-linearly realised, and as a non-linearly realised
transformation acts as a shift on certain scalar $\phi$,  $\d \phi
= \alpha + O(\phi)$, it cannot be preserved by any vacuum
expectation value of $\phi$ and so is necessarily broken by any
vacuum, so that the gauge group is necessarily broken down to its
linearly realised subgroup. Then any given vacuum solution (e.g. one
 arising from   a critical point of the scalar potential) can
then  break the linearly realised subgroup further to the subgroup
preserving that vacuum.

In this section, we will discuss the first stage of symmetry
breaking down to the linearly realised subgroup that is generic
for any solution. For vacua with vanishing scalar expectation
value, this is the complete breaking, but for non-trivial scalar
expectation values there will be further breaking through the
standard Higgs mechanism. The transformation for the scalar fields
$B_{(0)mn}$ is
\begin{equation}
\delta B_{(0)mn}=-\lambda_{(0)p}f^p_{mn}+K_{mnp}\omega^p
+2B_{(0)[m|p}f^p_{|n]q}\omega^q
\end{equation}
and from this one can find the non-linearly realised
symmetries, i.e. the ones realised as shifts of scalar fields.

Consider first the case without flux, $K_{mnp}=0$, so that the
gauge group with algebra (\ref{B-field eqiv. class algebra}) is $\cG _{2d}=\cG
_{d} \ltimes U(1)^d$ where $\cG _{d}$ is the isometry group with
structure constants $f^p_{mn}$. The transformation
\begin{equation}
\delta
B_{(0)mn}=-f^p_{mn}\lambda_{(0)p}+O(B_{(0)mn})\end{equation}
implies that  the subgroup of $Z$ transformations $\cG _{d}$ is
linearly realised, but that some of the $U(1)^d$ $X$
transformations with parameter $\lambda$ are non-linearly realised
and so broken. If $\cG _{d}$ is semi-simple, then one can raise
and lower group indices using the Cartan-Killing metric and define
the Goldstone field
\begin{equation}
\chi_p=\frac{1}{2}f^{mn}_pB_{(0)mn}
\end{equation}
which transforms as
\begin{equation}
\delta_X(\lambda_{(0)m})\chi_m=-\lambda_{(0)m}
\end{equation}
and the remaining scalar fields
\begin{equation}
\breve{B}_{(0)mn}=B_{(0)mn} - f^p_{mn}\chi_p
\end{equation}
transform linearly. Then the subgroup $U(1)^{d}$ of
$X$-transformations are non-linearly realised and so spontaneously
broken, with the Goldstone fields $\chi_p$ eaten by the gauge
fields $B_{(1)p}$ which all become massive. The group $\cG
_{2d}=\cG _{d} \ltimes U(1)^d$ is broken to the  subgroup $\cG
_{d}$, and this in turn may be further broken in vacua in  which
the  remaining scalars $\breve{B}_{(0)mn},g_{mn}$ have a
non-trivial vacuum expectation value. If the group is not
semi-simple, then not all of $U(1)^{d}$ is broken in general, and
the linearly realised subgroup
 is unbroken. In the trivial case in which $\cG _{d}$  is abelian, there are no shifts and the full
$U(1)^{2d}$ symmetry is unbroken, while in the general case it
will be broken to the subgroup for which the infinitesimal
parameters $\l$ satisfy
\begin{equation}
f^p_{mn}\lambda_{(0)p}=0
\end{equation}
For the abelian case $(f^m_{np}=0) $ the whole group remains
unbroken, for the semi-simple case there are no solutions and
$G_{2d}$ is broken to $G_{d}$, and for non-semi-simple cases there
in general will be solutions in the kernel of the map $\R^d \to
\R^{d(d-1)/2}$ defined by $\a_p \to \a_p f^p_{mn}$, resulting in a
partial breaking. This case will be analysed as a particular case
of a more general construction in section 6.5.2.

Consider next the case with flux, but with a toroidal reduction
with $f^p_{mn}=0$, so that $\cG _{d} = U(1)^d$. Then
\begin{equation}
\delta B_{(0)mn}=K_{mnp}\omega^p
\end{equation}
and  the $X$-transformations are linearly realised but a subgroup
of the group $\cG _{d} = U(1)^d$ of $Z$-transformations  is
non-linearly realised and so spontaneously broken. The breaking
depends on the form of the flux $K_{mnp}$. The $Z$ transformations
will be broken to the subset for which the parameters satisfy
$K_{mnp}\omega^p=0$. The remaining $Z$-transformations will be
broken, with   the corresponding scalars $ B_{(0)mn}$ eaten by
gauge fields $A^{m}$, which become massive. In this case, the
unbroken generators correspond  to the  kernel of the map $\R^d
\to \R^{d(d-1)/2}$ defined by $\b^p \to \b^p K_{mnp}$. This case
will be analysed in section 6.5.1.

We will now consider further examples of twisted reductions with
fluxes and discuss their symmetry breaking. In 6.2 we consider a
flux that may be removed by a simple field redefinition. Section
6.3 discusses a theory with a flux constructed from an invariant
metric, in which a linear combination of $B_{(1)m}$ and $A^m$
gauge fields becomes massive. Section 6.4 considers the general case and
finally, in section 6.5, we  illustrate the
general approach with two  examples.

\subsection{Trivial Flux}

The flux
\begin{equation}
K_{mnp}=\zeta_{mq}f^q_{np}+\zeta_{nq}f^q_{pm}+\zeta_{pq}f^q_{pm}
\end{equation}
 satisfies (\ref{eq:B-field integrability}) for any  constant
antisymmetric $\zeta_{nm}=-\zeta_{mn}$. The physical effect of
this flux, its appearance in the gauge algebra and in the
Lagrangian, may be removed entirely by the field redefinitions.
\begin{eqnarray}\label{trivial flux redefs}
\widetilde{B}_{(2)}&=&B_{(2)}+\frac{1}{2}\zeta_{mn}A^m\wedge A^n
\nonumber\\
\widetilde{B}_{(1)m}&=&B_{(1)m} - \zeta_{mn}A^n
\nonumber\\
\widetilde{B}_{(0)mn}&=&B_{(0)mn}+\zeta_{mn}
\end{eqnarray}
Fluxes of this form are therefore not physically significant.

\subsection{Flux Constructed from Invariant Metric }

If the group  ${\cal G}_d$ has an invariant metric $h_{mn}=h_{nm}$
satisfying
\begin{equation}
h_{(m|p}f^p_{|n)q}=0
\end{equation}
then
\begin{equation}
f_{mnp}= h_{mq}f^q_{np}
\end{equation}
is totally antisymmetric, $f_{mnp}=f_{[mnp]}$, and the flux
\begin{equation}
K_{mnp}=f_{mnp}
\end{equation}
satisfies the integrability constraint (\ref{eq:B-field
integrability}) by virtue of the Jacobi identity (\ref{eq:ff=0}).
If ${\cal G}_d$ is semi-simple then any invariant metric is
proportional to the Cartan-Killing metric $ \eta_{mn}$, so that
\begin{equation}
h_{mn}=\m \eta_{mn}, \qquad  \eta_{mn}=\frac{1}{2}f_{mp}^qf_{nq}^p
\end{equation}
for some parameter $\m$, which plays the role of a mass parameter
in the reduced theory. This metric is invertible, and is the case
considered for compact groups in \cite{Cvetic:2003jy}. The special
case $\m=0$ gives vanishing flux $K=0$ and so gives the reduction
for a semi-simple group without flux discussed in Section 6.1 as a
special case. For a general non-semi-simple  group,  $h_{mn}$ need
not be related to the Cartan-Killing metric and need not be
invertible.

We shall see that theories with this type of flux admit various
field redefinitions which simplify the gauge algebra of the
theory. The first step involves a change of basis  which brings
the gauge group to a semi-direct product form. If the metric $h_{mn}$ is invertible, then a further
redefinition is possible that brings the gauge group to a direct
product form ${\cal G}_d\times U(1)^d$ and this is spontaneously
broken to ${\cal G}_d$.

\subsubsection{General Case with Flux from Invariant Metric}

If $h_{mn}$ is an invariant metric, the flux $K_{mnp}=f_{mnp}$ is
a non-trivial flux in that its effect cannot be removed by a field
redefinition of the kind (\ref{trivial flux redefs}). The gauge
algebra   is (\ref{B-field eqiv. class algebra})
 with $K_{mnp}=f_{mnp}$
\begin{eqnarray}
\left[Z_m,Z_n\right]&=&f^p_{mn}Z_p-f_{mnp}X^p
\nonumber\\
\left[X^m,Z_n\right]&=&-f^m_{np}X^p
\nonumber\\
\left[X^m,X^n\right]&=&0
\end{eqnarray}
With this choice of flux, the algebra may be simplified by the
change of basis
\begin{equation}
\widehat{Z}_m=Z_m+h_{mn}X^n
\end{equation}
so that
\begin{equation}
\delta_{\widehat{Z}}(\omega^m)\equiv\delta_Z(\omega^m)+\delta_X(\lambda_{(0)m}\equiv
h_{mn}\omega^n)
\end{equation}
The gauge algebra  is then isomorphic to that of the standard
Scherk-Schwarz reduction in the absence of flux
\cite{Scherk:1979zr}
\begin{eqnarray}
[\widehat{Z}_m,\widehat{Z}_n]&=&f^p_{mn}\widehat{Z}_p
\nonumber\\
\left[  X^m,\widehat{Z}_n  \right]&=&-f^m_{np}X^p
\nonumber\\
\left[X^m,X^n\right]&=&0
\end{eqnarray}
which generates the gauge group
\begin{equation}
{{\cal G}}_d\ltimes U(1)^d
\end{equation}

The  $\widehat{Z}$ transformations  are
\begin{eqnarray}
\delta_{\widehat{Z}}(\omega)B_{(2)}&=&-h_{mn}\omega^mdA^n
\nonumber\\
\delta_{\widehat{Z}}(\omega)B_{(1)m}&=&h_{mn}d\omega^n+B_{(1)n}f^n_{mp}\omega^p
\nonumber\\
\delta_{\widehat{Z}}(\omega)B_{(0)mn}&=&B_{(0)mp}f^p_{nq}\omega^q-B_{(0)np}f^p_{mq}\omega^q
\nonumber\\
\delta_{\widehat{Z}}(\omega)A^m&=&-d\omega^m-f^m_{np}\omega^pA^n
\end{eqnarray}
Note that $\delta_{\widehat{Z}}(\omega)A^m=\delta_{Z}(\omega)A^m$
since the $A^m$ fields are singlets of the $X^m$ transformations.
The similarity of the $A^m$ and $B_{(1)m}$ transformations may be
exploited to define a field $C_{(1)m}$ which transforms
covariantly under the semi-simple subgroup $\widehat{{\cal G}}_d$
generated by $\delta_{\widehat{Z}}(\omega)$
\begin{equation}\label{C-field def}
C_{(1)m}=B_{(1)m}+h_{mn}A^n
\end{equation}
where
\begin{eqnarray}
\delta_{\widehat{Z}}(\omega)C_{(1)m}&=&C_{(1)n}f^n_{mp}\omega^p
%\nonumber\\
%\delta_Z(\omega)C_{(1)m}&=&C_{(1)n}f^n_{mp}\omega^p-h_{mn}D\omega^n
\nonumber\\
\delta_X(\lambda)C_{(1)m}&=&D\lambda_{(0)m}
\end{eqnarray}
The field strengths (\ref{G field strengths}) then become
\begin{eqnarray}\label{s-simple field strengths}
G_{(3)}&=&dB_{(2)}+C_{(1)m}\wedge F^m - \Omega_3(h)
\nonumber\\
G_{(2)m}&=&DC_{(1)m}+B_{(0)mn}F^n-h_{mn}F^n
\nonumber\\
G_{(1)mn}&=&DB_{(0)mn}+f^p_{mn}C_{(1)p}
\nonumber\\
G_{(0)mnp}&=&- 3B_{(0)[m|q}f^q_{|np]}+f_{mnp}
\end{eqnarray}
where
\begin{equation}
\Omega_3(h)=h_{mn}\left(A^m\wedge
dA^n+\frac{1}{3}f^m_{pq}A^n\wedge A^p\wedge A^q \right)
\end{equation}
is a generalised Chern-Simons term satisfying
\begin{equation}
d\Omega_3(h) =h_{mn} F^m\wedge F^n
\end{equation}
These field redefinitions
 give a formulation
 with gauge fields
$(C_{(1)m},A^m)$ and gauge group $ {{\cal G}}_d\ltimes U(1)^d$
with $A^m$ the gauge fields for the $\widehat{Z}$ transformations
generating $\cG_d$ with the $U(1)^d$  gauge fields
  $C_{(1)m}$   transforming covariantly under $\cG_d$.
 The   field redefinition (\ref{C-field def}) brings the vector mass term in
  the Lagrangian to the form
\begin{eqnarray}
{\cal L}_D&=&-\frac{1}{2}e^{-\phi}g^{mn}g^{pq}*G_{(1)mp}\wedge
G_{(1)nq}+...\nonumber\\
&=&-\frac{1}{2}e^{-\phi}g^{mn}g^{pq}f^t_{mp}f^s_{nq}*C_{(1)t}\wedge
C_{(1)s}+...
\end{eqnarray}

so that only the $C_{(1)m}$ can become massive (which ones do so
depends on the structure constants) while the $A^m$ remain
massless.

\subsubsection{Case with Invertible Metric}

To go further, we now restrict ourselves to the case in which
$h_{mn}$ is non-degenerate, with inverse $h^{mn}$. In this case
the gauge group may be simplified further. An important class of
examples is that in which
  $\cG_d$ is  a semi-simple group, and the
invariant metric $h_{mn}=\m \eta_{mn}$ is proportional to the
Cartan-Killing metric $\eta_{mn}$. Then further redefinitions are
possible that simplify the gauge group   to a direct product $
{{\cal G}}_d\times U(1)^d$ which is spontaneously broken to $
{{\cal G}}_d$ with  the gauge fields $C_{(1)m}$ becoming massive
\cite{Cvetic:2003jy}.

The gauge fields $C_{(1)m}$ become massive by eating the Goldstone
fields
\begin{equation}
\chi_p=\frac{1}{2}f^{mn}_pB_{(0)mn}
\end{equation}
where $f^{mn}_p=h^{mq}f^n_{qp}$. This transforms as
\begin{eqnarray}
\label{eq:B-field stuckelberg transformations}
\delta_{\widehat{Z}}(\omega^m)\chi_{m}&=&\chi_nf^n_{mp}\omega^p
\nonumber\\
%\delta_Z(\omega^m)\chi_{m}&=&\chi_nf^n_{mp}\omega^p+\eta_{mn}\omega^n
%\nonumber\\
\delta_X(\lambda_{(0)m})\chi_m&=&- \lambda_{(0)m}
\end{eqnarray}
so that the $\chi$ are Stuckelberg fields shifting under the $X$
transformations (if $\m\ne0$).

The gauge algebra can be further simplified to a direct product
structure by defining massive fields which are gauge singlets
under $\delta_X(\lambda_{(0)m})$, as follows
\begin{eqnarray}\label{massive B-field def}
\breve{B}_{(2)}&=&B_{(2)} - \chi_mF^m
\nonumber\\
\breve{C}_{(1)m}&=&C_{(1)m} + D\chi_m
\nonumber\\
\breve{B}_{(0)mn}&=&B_{(0)mn} - f^p_{mn}\chi_p
\end{eqnarray}
The
 Goldstone bosons  $\chi_{m}$ of the broken $U(1)^d$
symmetry $\delta_X(\lambda_{(0)m})$ are eaten by the $C_{(1)m}$
which  become the massive vector fields $\breve{C}_{(1)m}$. These
redefinitions bring the ${\widehat{Z}} $ transformations for
$\breve{C}_{(1)m}$ and $\breve{B}_{(0)mn}$ to the canonical form,
\begin{eqnarray}\label{massive B-field diffeos}
\delta_{\widehat{Z}}(\omega^m)\breve{B}_{(2)}&=&-h_{mn}\omega^mdA^n \nonumber\\
\delta_{\widehat{Z}}(\omega^m)\breve{C}_{(1)m}&=&\breve{C}_{(1)n}f^n_{mp}\omega^p\nonumber\\
\delta_{\widehat{Z}}(\omega^m)\breve{B}_{(0)mn}&=&\breve{B}_{(0)mp}f^p_{nq}\omega^q-\breve{B}_{(0)np}f^p_{mq}\omega^q
\end{eqnarray}

The gauge algebra realised on     $\breve{B}_{(0)mn}$ and
$\breve{C}_{(1)m}$ is
\begin{equation}
\lbrack
\delta_{\widehat{Z}}(\widetilde{\omega}),\delta_{\widehat{Z}}(\omega)
\rbrack =
\delta_{\widehat{Z}}(f^m_{np}\omega^n\widetilde{\omega}^p)
\end{equation}
Since the massive potentials are singlets of
$\delta_X(\lambda_{(0)m})$ this is the complete algebra. The gauge
algebra is then
\begin{eqnarray}
\left[\widehat{Z}_m,\widehat{Z}_n\right]&=&f^p_{mn}\widehat{Z}_p
\nonumber\\
\left[X^m,\widehat{Z}_n\right]&=&0
\nonumber\\
\left[X^m,X^n\right]&=&0
\end{eqnarray}
which generates the gauge group
\begin{equation}
{\cal G}_d\times U(1)^d
\end{equation}
The action of $\delta_{\widehat{Z}}(\omega)$ on the
$\breve{B}_{(0)mn}$ and $\breve{C}_{(1)m}$ fields is equivalent to
that of $\delta_{Z}(\omega)$ since these fields are singlets of
the $U(1)^d$ subgroup. The $\breve{B}_{(2)}$ field is still
massless and transforms as
$\delta_W(\Lambda_{(1)})\breve{B}_{(2)}=d\Lambda_{(1)}$. The full
gauge algebra is
\begin{equation}
\lbrack
\delta_{\widehat{Z}}(\widetilde{\omega}),\delta_{\widehat{Z}}(\omega)
\rbrack =
\delta_{\widehat{Z}}(f^m_{np}\omega^n\widetilde{\omega}^p)-\delta_W(f_{mnp}\omega^n\widetilde{\omega}^pA^m)
\end{equation}
with all other commutators vanishing.

The field redefinitions (\ref{massive B-field def}) are of the
form of the infinitesimal gauge transformations generated by
$\delta_X(\lambda)$, although they are not infinitesimal. Since
the field strengths are invariant under transformations generated
by $\delta_X(\lambda)$ the field strengths take the same form as
(\ref{s-simple field strengths}), with $B_{(2)}$, $C_{(1)m}$ and
$B_{(0)mn}$ replaced by $\breve{B}_{(2)}$, $\breve{C}_{(1)m}$ and
$\breve{B}_{(0)mn}$ respectively.

\subsection{General Case}

In this section we shall consider a reduction on a general twisted
torus with arbitrary flux, i.e. we shall allow the
$d$-dimensional group ${\cal G}_d$, upon which the
compactification is based, to be non-semi-simple and the flux $K$
may vanish along certain directions. We shall  analyse
  the symmetry breaking of the gauge symmetry ${\cal
G}_{2d}\simeq {\cal G}_d\ltimes U(1)^d$ to a linearly realised
subgroup.

The gauge fields $A^m, B_{(1)m}$ become massive by eating a subset
of the scalar fields $B_{(0)mn}$, so it is on these fields that we
shall focus. The infinitesimal transformation of the scalar
$B$-fields under the gauge symmetry ${\cal G}_{2d}$ is
\begin{equation}
\delta B_{(0)M}=-\lambda_{(0)m}f^m{}_M+\omega^mK_{mM}+O(B_{(0)M})
\end{equation}
where we have defined the compound index $M=[mn]$, $M=1,2,...D$,
where $D=d(d-1)/2$. It will be useful to write this variation in
terms of the $O(d,d)$ covariant structure constants $t_{AB}{}^C$
of ${\cal G}_{2d}$ and the gauge parameter
$\alpha_A=L_{AB}\alpha^B$ introduced in section 5.3 where
$A=1,2,...2d$.
\begin{eqnarray}
\delta B_{(0)M}&=&\left( -\lambda_m, \omega^m\right)
\left(\begin{array}{cc} f^m{}_M \\
 K_{mM}
\end{array}\right)+O(B_{(0)M})\nonumber\\
&&=\alpha_At^A{}_M+O(B_{(0)M})
\end{eqnarray}
The structure constants $t^A{}_M$ may be thought of as defining a map
$\alpha _A \to \alpha _A t^A{}_M$
 from
$\R^{2d}$ to $\R^D$, which is non-invertible for $d>5$ (and may be for $d\le 5$), and   part of our analysis will be concerned with finding
a generalised inverse of this map.

First we identify the set of gauge fields which become massive.
The map $t: \alpha _A \to \alpha _A t^A{}_M$ will have a kernel of dimension $2d-d'$ for some $d'$, and it is useful to choose a basis
$\{ e_A \}=\{ e_{A'}, e_{\bar {A}} \} $ for $R^{2d}$ consisting of
a basis
$\{ e_{\bar{A} }\}$ for the  kernel of the map $t$ with $\bar{A}=d'+1,...2d$
together with its complement, a basis $\{e_{A'}\} $ for the cokernel, with $A'=1,2...d'$.
Then \begin{equation}\label{newT2}
t^{\bar{A}}{}_M= 0
\end{equation}
for \emph{all} $M$
while $t^{A'}{}_M$ has no zero eigenvectors and so for each $A'$, there is \emph{some} $M$ such that $t^{A'}{}_M\neq 0$.
Then in this basis   the
structure constants $t^A{}_M$ are a $2d\times D$ matrix
\begin{equation}\label{newT}
t^A{}_M=\left(\begin{array}{cc} t^{A'}{}_M \\
 0
\end{array}\right)
\end{equation}
where $0$ is the $2d-d'\times D$ zero matrix.

In this  basis the $B_{(0)M}$ scalar
field transforms as
\begin{equation}
\delta B_{(0)M}=\alpha_{A'}t^{A'}{}_M+O(B_{(0)M})
\end{equation}
so that the symmetry ${\cal G}_{2d}$ is broken to the $2d-d'$
dimensional subgroup generated by the $T^{\bar{A}}$ with
parameters $\alpha_{\bar{A}}$ and gauge bosons ${\cal
A}_{\bar{A}}$, while the remaining symmetries generated by $T^{
{A'}}$ are all broken and the vector fields ${\cal A}_{A'}$ are
massive.
 Indeed, the mass term of the Lagrangian \ref{Odd Lagrangian} for the
gauge fields ${\cal A}_A$ is
\begin{equation}
{\cal L}_D=\frac{1}{4}e^{-\phi}{\cal M}^{AB}{\cal
M}^{CD}t_{AC}{}^{E'}t_{BD}{}^{F'}*{\cal A}_{E'}\wedge {\cal
A}_{F'}+...
\end{equation}
and it is clear that fields ${\cal A}_{A'}$ are massive whilst
the ${\cal A}_{\bar{A}}$ remain massless, as the latter do not
appear in the mass term.
 In general, $A_{A'}$ and ${\cal A}_{\bar{A}}$ will be
linear combinations of the $A^m$ and $B_{(1)m}$ fields.
 This highlights one effect of introducing
flux into such twisted reductions: if $K=0$, then only a subset of
the $B_{(1)m}$ fields become massive whilst the $A^m$ remain
massless. By introducing fluxes, the fields
which become massive are linear combinations of the $A^m$ and $B_{(1)m}$ fields.

We now turn to finding the Goldstone fields that are eaten by the
massive one-form fields ${\cal A}_{A'}$.
 The structure constants also give a map from $\R^D $ to $\R^{2d}$ defined by $\beta ^M \to t^A{}_M \beta ^M $. As before we choose a basis for
 $\R^D $ so that the index $M$
 splits
into $(M',\bar{M})$ where $\bar{M}=d'+1,..D$ label a basis for the kernel of this map, and  $M'=1,2,..d'$ labels a basis for  the cokernel.
Then $t^A{}_M$ takes the form of a $2d\times D$ matrix
\begin{equation}
t^A{}_M=\left(\begin{array}{cc} t^{A'}{}_{M'} & 0
\\
 0 & 0 \end{array}\right)
\end{equation}
The $2d\times D$ matrix $t^A{}_M$ is non-invertible, but  the $d'\times d'$ matrix $t^{A'}{}_{M'}$ is non-degenerate by construction, so
we may
define its inverse
$\tilde{t}^{M'}{}_{A'}$ (with $\tilde{t}^{M'}{}_{A'}t^{A'}{}_{N'}=\delta
^{M'}{} _{N'}$).

Now
\begin{equation}
 \delta
B_{(0)\bar{M}}=O(B_{(0) })
\qquad  \delta
B_{(0) {M'}}=\alpha _{A'}t^{A'}{}_{M'}
+O(B_{(0) })
\end{equation}
so that we can define
Goldstone bosons $\chi_{A'}$ by $\chi_{A'}=B_{(0)M'}\tilde{t}^{M'}{}_{A'}$
so that they transform as
\begin{equation}
\delta\chi_{A'}=\alpha_{A'}+O(B_{(0) })
\end{equation}
We may then define
massive gauge bosons as
\begin{equation}
\breve{{\cal A}}_{A'}={\cal A}_{A'}+D\chi_{A'}
\end{equation}
so that the massive gauge fields $\breve{{\cal A}}_{A'}$ transform
in a linear realisation of the unbroken gauge group generated by
$T^{\bar{A}}$.

\subsection{Examples}

In the following we illustrate this general method with two
examples.

\subsubsection{Toroidal Reduction with Flux}

If $f^p_{mn}=0$, then the group ${\cal G}_d$ is abelian and the
internal manifold (after discrete identifications to compactify,
if necessary) is   a torus and we take $\cG _{d} = U(1)^d$. With
flux $K$, the gauge algebra  is
\begin{eqnarray}
\left[Z_m,Z_n\right]&=&-K_{mnp}X^p
\end{eqnarray}
with all other commutators vanishing. The internal index $m$ can
be split into $(m',\bar{m})$, so that
 $\bar{m}$ labels the $d-d'$ dimensional kernel of
 the map
 $\alpha^m \to \alpha^m K_{mnp}$, and $m'$ labels the cokernel, so that
\begin{equation}
K_{mn\bar{p}} =0 \qquad
K_{m'n'p'} \neq 0
\end{equation}
Then the   transformation of the
$B_{(0)}$ scalars is
\begin{equation}
\delta B_{(0){ n'p'}} =\omega^{m'}K_{m'n'p'}
, \qquad \delta B_{(0){ m\bar{n}}} = 0
\end{equation}
The transformations generated by $Z_{m'}$ with parameters
$\omega^{m'}$ are spontaneously broken, with $B_{(0)m'n'}$ the
Goldstone fields that are eaten by the gauge fields $A^{m'}$. The
$A^{m'}$ fields have mass term in the Lagrangian
\begin{equation}
\mathcal{L}_D=-\frac{1}{2}e^{-\phi}g^{mn}g^{pq}K_{mpt'}K_{nqs'}*A^{t'}\wedge
A^{s'}+...
\end{equation}
The $2d$ dimensional gauge group is broken to the $2d-d'$
dimensional abelian subgroup $U(1)^{2d-d'}$ generated by $Z_{\bar
m}$ and $X^m$ with parameters $\omega^{\bar{m}}$ and $\lambda_m$
respectively.

Let $\tilde K^{m'n'p'}$  be any constants satisfying
$\tilde K^{m'n'p'}K_{ n'p'q'}=\delta ^{m'}{}_{q'}$. Then
we can define Goldstone fields $\chi^{m'}$
by
\begin{equation}
    \chi^{m'}=\tilde K^{m'n'p'}B_{n'p'}
\end{equation}
transforming as a shift the $\omega^{m'}$
transformations
\begin{equation}
\delta B_{(0)\bar{M}}=0 \qquad \delta\chi^{m'}=\omega^{m'}
\end{equation}
The remaining scalars are invariant, $\delta B_{(0){ m\bar{n}}} = 0$.
We may then define the massive graviphotons
$\breve{A}^{m'}=A^{m'}+d\chi^{m'}$ which are singlets of the gauge
transformations.

\subsubsection{Reduction with Non-semi-simple Twisted Tori}

The next example we consider is that of a general non-semi-simple
group $\cG_d$  with structure constants $f^p_{mn}$ but with zero
flux, so that the analysis of subsection 6.3.1 applies with
$\m=0$. Such groups arise for example in reductions with duality
twists. The reduction is then on a compactification $\cG_d/\Gamma$
of the non-compact group manifold  for the non-semi-simple group
$\cG_d$.

The map $\R^{d} \to \R^{d(d-1)/2}$ defined by $V_{m} \to V_{m}
f^{m}_{np}$ will have a kernel of dimension $d-d'$, say, which we
label by $\bar{m}=d'+1,...,d$, while the cokernel is labelled by
$m'=1,..., d'$, so that the index $m$ is split
 into $(m',\bar{m})$. Then
\begin{equation}
f^{\bar{m}}{}_{np}=0    \qquad \forall n,p
\end{equation}
and the map
 can be written as $V_{m'} \to V_{m'} f{}^{m'}{}_{np}$.

The transformation of the $B_{(0)mn}$ scalars under ${\cal G}_{2d}$
is
\begin{equation}\label{B transformation}
\delta B_{(0)np}=-\lambda_{m'}f^{m'}{}_{np}+O(B_{(0) })
\end{equation}
so that the
transformations generated by the $X^{m'}$ are broken, with the
vector fields
 $B_{(1)m'}$ becoming massive. Indeed,  the fields $B_{(1)m'}$ have a   mass
 term in the Lagrangian
\begin{equation}
{\cal L}_D=
-\frac{1}{2}e^{-\phi}g^{mn}g^{pq}f^{t'}_{mp}f^{s'}_{nq}*B_{(1)t'}\wedge
B_{(1)s'}+...
\end{equation}
 The gauge group is then broken to the $2d-d'$ dimensional subgroup generated by
 $Z_m$ and $X^{\bar{m}}$ with parameters $\omega^m$ and $\lambda_{\bar{m}}$, with massless gauge fields $A^m, B_{\bar{m}}$.

To proceed with the analysis, we define a compound index $M=[mn]$
  $M=1,..,D$, where $D=d(d-1)/2$, labelling the space of
2-forms $ \R^D$, so that the structure constants define a $d\times
D$ matrix $f^m{}_N$. We then split this into indices
$M=(M',\bar{M})$, $M'=1,2,...d'$, $\bar{M}=d'+1,...D'$ with $M'$
labelling the cokernel of the map defined by $\beta ^M \to
f^m{}_M\beta ^M$ and $\bar{M}=d'+1,...D'$ labelling the kernel.
Then $f^{m'}{}_{M'}$ is an invertible $d'\times d'$ matrix with
inverse $\tilde f ^{M'}{}_{m'}$, say.
 The goldstone
bosons of the broken symmetry $\chi_{m'}$
are $\chi_{m'}=B_{(0)M'}\tilde f^{M'}{}_{m'}$
 and the scalar fields
transforming linearly in the unbroken gauge symmetry ${\cal
G}_d\ltimes U(1)^{d-d'}$ are
  $B_{(0)\bar{M}}$.
  The transformation properties of these fields under
$X$ and $Z$ transformations following   from (\ref{B
transformation})
 are
\begin{equation}
\delta B_{(0)\bar{M}}=O(B_{(0)}) \qquad
\delta\chi_{m'}=\lambda_{m'}+O(B_{(0)})
\end{equation}
so that $\chi_{m'}$ is the Goldstone field for the $\lambda_{m'}$
transformations. We   then define the massive gauge boson
$\breve{B}_{(1)m'}=B_{(1)m'}+D\chi_{m'}$ which transforms in a
linear representation of the unbroken gauge group ${\cal
G}_d\ltimes U(1)^{d-d'}$.

As an example, consider the case where $d=3$  ,
$m' =1,2$ and $\bar {m}=3$, with the only non-zero structure constants given by
$f^{m'}_{3n'}=M _{n'}{}^{m'}$ for some matrix $M _{n'}{}^{m'}$, so
that the gauge algebra, in the absence of flux, becomes
\begin{eqnarray}
\left[Z_{m'},Z_i\right] &=&M _{m'}{}^{n'}Z_{n'}
\nonumber\\
\left[X^{m'},Z_i\right] &=&-M _{n'}{}^{m'}X^{n'}
\end{eqnarray}
with all other commutators vanishing. This is precisely the
algebra (\ref{duality twist algebra}) of an $SL(2)$ twisted $T^2$
fibration over $S^1$ as discussed in section 3, with mass matrix
$M_{m'}{}^{n'}$. For example, in the case of an elliptic twist,
the gauge group $ISO(2)\ltimes U(1)^3$ is broken to the linearly
realised sub-group $ISO(2)\ltimes U(1)$.

%%%%%%%%%

\section{Discussion}

We have seen that a Scherk-Schwarz reduction of a low energy field
theory on a \lq twisted torus' associated with a group $\cG$ can
be extended to a string theory compactification provided the group
$\cG$ has a discrete cocompact subgroup $\G$, and the string
theory is then compactified on $\cG/\G$. This is a non-trivial
restriction, as not all $\cG$ have such subgroups. A large class
(but not all) of reductions with duality twists can be viewed as
compactifications on such $\cG/\G$, while others can be regarded as
compactifications of F-thoery, or one of its generalisations, on
$\cG/\G$. This extends to M-theory: a
Scherk-Schwarz reduction of 11-dimensional supergravity can be
promoted to a compactification of M-theory on a compact $\cG/\G$,
when this exists.

The $O(d,d)$ covariant formulation of the reduced theory is very
suggestive, and it is natural to ask whether this generalises to
M-theory or type II compactifications, and whether these can be
written in a way that is covariant under the action of a duality
group.
 The results for the heterotic string with Wilson line fluxes for the
heterotic gauge fields were given in \cite{Odd}, and were found to
be $O(d,d+16)$ covariant. The results are of the form given in
section 5, but with the indices $A$ now running over $2d+16$
values and transforming under $O(d,d+16)$.

For general Scherk-Schwarz  compactifications of string theory,
the vector fields $B_{\m m}$ arise from the
2-form gauge field $B_{MN}$. These
correspond to  the gauge group generators $X^m$ and couple to string winding modes. In generalising to M-theory,
these are replaced by vector fields $C_{\m mn}$ ($C_{\m mn}=-C_{\m nm}$)
arising from the
3-form gauge field  $C_{MNP}$. These are  associated with
group generators $X^{mn}=- X^{nm}$ and couple to
membrane wrapping modes.
This gives a gauge group whose generators include $Z_m, X^{mn}$, with algebra
\begin{eqnarray}
\left[Z_m,Z_n\right]&=&f^p_{mn}Z_p-K_{mnpq}X^{pq}
\nonumber\\
\left[X^{mn},Z_p\right]&=&2f^{[m}_{pq}X^{n]q}
\nonumber\\
\left[X^{mn},X^{pq}\right]&=&0
\end{eqnarray}
where $K_{mnpq}$ are the constants defining the 4-form flux. This
algebra was also found by \cite{Dall'Agata:2005ff} for
Scherk-Schwarz reductions of 11-dimensional supergravity to 4
dimensions. One might have expected that the $O(d,d)$ covariance
reviewed here would extend to a covariance under the appropriate
U-duality group. However, we find that in general fluxes provide
obstructions to the dualisations needed to bring the theory to a
more symmetric form with U-duality covariance, but nonetheless, an
elegant and suggestive structure emerges, with covariance under
the \lq electric subgroup' of the U-duality group (i.e. the
subgroup that can be realised on the gauge potentials). The
details will be given in \cite{M-Theory paper}.

The toroidal reduction of the field theory action gives a theory
with a rigid $O(d,d)$ duality symmetry, and the twisted torus
reduction with flux gives a theory in which a $2d$ dimensional
subgroup $\cG_{2d}$ of $O(d,d)$ is promoted to a gauge group. The
gauge fields are in the vector representation of  $O(d,d)$, so
this must become the adjoint of $\cG_{2d}\subset O(d,d)$. If the
original theory is a supergravity theory, the result is a gauged
supergravity theory. Consider now the lifting of this to string
theory. The $O(d,d)$ duality symmetry of the toroidal reduction is
broken to the T-duality group $O(d,d;\Z)$ yet a subgroup of the
continuous group $O(d,d)$ is meant to be a gauge symmetry. This
raises the issue of how the two symmetries are related. This was
addressed in \cite{Odd}, where it was  suggested that they were distinct and
that there is  an $O(d,d;\Z)\times \cG_{2d}$ gauge symmetry.
However, we will see that the situation is rather subtle, and that
although they are distinct, they do not commute.

The structure constants $t_{AB}{}^C$ specify the embedding of
$\cG_{2d}\subset O(d,d)$. In the field theory reduction of section
5, there is covariance under $O(d,d)$ which acts non-trivially on
the structure constants $t_{AB}{}^C$, as they transform
covariantly. Then an $O(d,d)$ transformation acts not just on the
fields, but on the coupling constants $t_{AB}{}^C$ and hence
changes the form of the mass terms and potential. The gauge group
is still $\cG_{2d}$, but after the transformation it is embedded
differently in $O(d,d)$. The new form of the theory is related to
the original one by field redefinitions, so it is physically
equivalent. A similar situation was discussed in
\cite{Hull:1983yf,Hull:1984qz,Hull:1984yy}, where the action of
$SL(8,\R)\subset E_7$ duality symmetries was considered on
 $N=8$ gauged supergravity in $D=4$, giving equivalent gaugings (although singular limits gave new gaugings).
Thus there is still an action of the rigid $O(d,d)$, but it does not leave the gauged theory invariant, but changes it to a physically equivalent theory, with
the gauge group $\cG_{2d}$ transformed to a conjugate gauge group $\cG_{2d}$ embedded differently in $O(d,d)$.

In the compactifications of string theory discussed here,
there is a  $\cG_{2d}$ gauge symmetry, but one might expect that the
$O(d,d)$ covariance should be broken to
a discrete subgroup.
A simple case in which this should happen is that of toroidal reductions with $H$-flux.
The $O(d,d)$ contains a geometric subgroup
$GL(d,\R)\ltimes \R^{d(d-2)/2}$,
 acting through diffeomorphisms and shifts of the $B$ field, and these are broken to
$GL(d,\Z)\ltimes \Z^{d(d-2)/2}$ by the torus boundary conditions
and the requirement that the quantum theory be invariant under
$B$-shifts. Then the $O(d,d)$ covariance should be broken to a
subgroup containing $GL(d,\Z)\ltimes \Z^{d(d-2)/2}$, so that a
natural conjecture would be that the theory should have
$O(d,d;\Z)$ covariance, and this is precisely what the low-energy
field theory suggests. However, the status of the T-duality
transformations in $O(d,d;\Z)$ is not clear. The transformation
corresponding to T-duality in one torus direction removes the flux
and turns on a twist, giving some structure constants $f^m_{np}$
and a twisted torus \cite{Hull:1998vy,Lowe:2003qy}. However, a
further T-duality would then take  the background to a
non-geometric background \cite{Hull:2004in} and so an
understanding of whether the conjecture that there should be
$O(d,d;\Z)$ covariance would require a generalisation of the
compactifications considered here to non-geometric backgrounds, and a
better understanding
of T-duality in such cases in which the usual rules do not apply.
However, our analysis makes it clear that if there is such a covariance under
$O(d,d;\Z)$ or some other discrete subgroup of $O(d,d)$,
 it would not be a conventional symmetry
acting on the fields alone, but must be duality that acts on the
coupling constants $t_{AB}{}^C$ as well, just as S-duality in
$N=4$ Yang-Mills acts on the coupling constant as well as the
fields. The flux and twist then transform into each other and
fit together into an irreducible representation. In particular,
the $O(d,d)$ action does not commute with the gauge symmetry.
Similar remarks should apply to the interplay between gauge symmetry and U-duality in
compactifications  of M-theory
, as we will discuss elsewhere.


\newpage

\begin{thebibliography}{03}

%\cite{Scherk:1978ta}
%\bibitem{Scherk:1978ta}
%J.~Scherk and J.~H.~Schwarz, ``Spontaneous Breaking Of
%Supersymmetry Through Dimensional Reduction,'' Phys.\ Lett.\ B
%{\bf 82}, 60 (1979).
%%CITATION = PHLTA,B82,60;%%



%\cite{Scherk:1979zr}
\bibitem{Scherk:1979zr}
J.~Scherk and J.~H.~Schwarz, ``How To Get Masses From Extra
Dimensions,'' Nucl.\ Phys.\ B {\bf 153}, 61 (1979).
%%CITATION = NUPHA,B153,61;%%



%\cite{Dabholkar:2002sy}
\bibitem{Dabholkar:2002sy}
A.~Dabholkar and C.~Hull, ``Duality twists, orbifolds, and
fluxes,'' JHEP {\bf 0309}, 054 (2003) [arXiv:hep-th/0210209].
%%CITATION = HEP-TH 0210209;%%


%\cite{Duff:1984hn}
\bibitem{Duff:1984hn}
M.~J.~Duff, B.~E.~W.~Nilsson, C.~N.~Pope and N.~P.~Warner, ``On
The Consistency Of The Kaluza-Klein Ansatz,'' Phys.\ Lett.\ B {\bf
149}, 90 (1984).
%%CITATION = PHLTA,B149,90;%%


%\cite{Duff:hr}
\bibitem{Duff:hr}
M.~J.~Duff, B.~E.~W.~Nilsson and C.~N.~Pope, ``Kaluza-Klein
Supergravity,'' Phys.\ Rept.\  {\bf 130}, 1 (1986).
%%CITATION = PRPLC,130,1;%%

%\cite{Cvetic:2000dm}
\bibitem{Cvetic:2000dm}
M.~Cvetic, H.~Lu and C.~N.~Pope, ``Consistent Kaluza-Klein sphere
reductions,'' Phys.\ Rev.\ D {\bf 62}, 064028 (2000)
[arXiv:hep-th/0003286].
%%CITATION = HEP-TH 0003286;%%


%\cite{Cvetic:2003jy}
\bibitem{Cvetic:2003jy}
M.~Cvetic, G.~W.~Gibbons, H.~Lu and C.~N.~Pope, ``Consistent group
and coset reductions of the bosonic string,'' Class.\ Quant.\
Grav.\  {\bf 20}, 5161 (2003) [arXiv:hep-th/0306043].
%%CITATION = HEP-TH 0306043;%%

%\cite{Hull:1988jw}
\bibitem{Hull:1988jw}
C.~M.~Hull and N.~P.~Warner, ``Noncompact Gaugings From Higher
Dimensions,'' Class.\ Quant.\ Grav.\  {\bf 5}, 1517 (1988).
%%CITATION = CQGRD,5,1517;%%



%\cite{Nicolai:1984jg}
\bibitem{Nicolai:1984jg}
H.~Nicolai and C.~Wetterich,
%``On The Spectrum Of Kaluza-Klein Theories With Noncompact Internal Spaces,''
Phys.\ Lett.\ B {\bf 150}, 347 (1985).
%%CITATION = PHLTA,B150,347;%%

%\cite{Gibbons:2001wy}
\bibitem{Gibbons:2001wy}
G.~W.~Gibbons and C.~M.~Hull,
%``de Sitter space from warped supergravity solutions,''
arXiv:hep-th/0111072.
%%CITATION = HEP-TH 0111072;%%


%\cite{Hull:2001ii}
\bibitem{Hull:2001ii}
C.~M.~Hull,
%``De Sitter space in supergravity and M theory,''
JHEP {\bf 0111}, 012 (2001) [arXiv:hep-th/0109213].
%%CITATION = HEP-TH 0109213;%%


%\cite{Lavrinenko:1997qa}
\bibitem{Lavrinenko:1997qa}
I.~V.~Lavrinenko, H.~Lu and C.~N.~Pope, ``Fibre bundles and
generalised dimensional reductions,'' Class.\ Quant.\ Grav.\  {\bf
15}, 2239 (1998) [arXiv:hep-th/9710243].
%%CITATION = HEP-TH 9710243;%%


%\cite{Kaloper:1998kr}
\bibitem{Kaloper:1998kr}
N.~Kaloper, R.~R.~Khuri and R.~C.~Myers, ``On generalized axion
reductions,'' Phys.\ Lett.\ B {\bf 428}, 297 (1998)
[arXiv:hep-th/9803066].
%%CITATION = HEP-TH 9803066;%%





%\cite{Bergshoeff:1997mg}
\bibitem{Bergshoeff:1997mg}
E.~Bergshoeff, M.~de Roo and E.~Eyras, ``Gauged supergravity from
dimensional reduction,'' Phys.\ Lett.\ B {\bf 413}, 70 (1997)
[arXiv:hep-th/9707130].
%%CITATION = HEP-TH 9707130;%%


%\cite{Bergshoeff:1996ui}
\bibitem{Bergshoeff:1996ui}
E.~Bergshoeff, M.~de Roo, M.~B.~Green, G.~Papadopoulos and
P.~K.~Townsend, ``Duality of Type II 7-branes and 8-branes,''
Nucl.\ Phys.\ B {\bf 470}, 113 (1996) [arXiv:hep-th/9601150].
%%CITATION = HEP-TH 9601150;%%

%\cite{Kaloper:1999yr}
\bibitem{Odd}
N.~Kaloper and R.~C.~Myers, ``The O(dd) story of massive
supergravity,'' JHEP {\bf 9905}, 010 (1999)
[arXiv:hep-th/9901045].
%%CITATION = HEP-TH 9901045;%%


%\cite{Hull:2002wg}
\bibitem{Hull:2002wg}
C.~M.~Hull, ``Gauged D = 9 supergravities and Scherk-Schwarz
reduction,'' arXiv:hep-th/0203146.
%%CITATION = HEP-TH 0203146;%%



%\cite{Bergshoeff:2002mb}
\bibitem{Bergshoeff:2002mb}
E.~Bergshoeff, U.~Gran and D.~Roest, ``Type IIB seven-brane
solutions from nine-dimensional domain walls,'' Class.\ Quant.\
Grav.\  {\bf 19}, 4207 (2002) [arXiv:hep-th/0203202].
%%CITATION = HEP-TH 0203202;%%


%\cite{Andrianopoli:2004xu}
\bibitem{Andrianopoli:2004xu}
L.~Andrianopoli, S.~Ferrara and M.~A.~Lledo,
%``No-scale D = 5 supergravity from Scherk-Schwarz reduction of D = 6
%theories,''
JHEP {\bf 0406}, 018 (2004)
[arXiv:hep-th/0406018].
%%CITATION = HEP-TH 0406018;%%



%\cite{Andrianopoli:2004im}
\bibitem{Andrianopoli:2004im}
L.~Andrianopoli, S.~Ferrara and M.~A.~Lledo, ``Scherk-Schwarz
reduction of D = 5 special and quaternionic geometry,'' Class.\
Quant.\ Grav.\  {\bf 21}, 4677 (2004) [arXiv:hep-th/0405164].
%%CITATION = HEP-TH 0405164;%%



%\cite{Andrianopoli:2004fc}
\bibitem{Andrianopoli:2004fc}
L.~Andrianopoli, S.~Ferrara and M.~A.~Lledo', ``Generalized
dimensional reduction of supergravity with eight supercharges,''
arXiv:hep-th/0410005.
%%CITATION = HEP-TH 0410005;%%



%\cite{D'Auria:2004td}
\bibitem{D'Auria:2004td}
R.~D'Auria, S.~Ferrara and M.~Trigiante, ``No-scale supergravity
from higher dimensions,'' arXiv:hep-th/0409184.
%%CITATION = HEP-TH 0409184;%%


%cite{cow}
\bibitem{cow} P.M. Cowdall, H. Lu, C.N. Pope, K.S. Stelle and
P.K. Townsend, ``Domain Walls in Massive Supergravities,'' Nucl.
Phys. {\bf B486} (1997) 49 [hep-th/9608173].
%%CITATION = HEP-TH 9608173

\bibitem{flux} I.V. Lavrinenko, H. Lu and  C.N. Pope, ``From Topolgy to Generalised Dimensional Reduction," Nucl. Phys. {\bf B492}
(1997) 278 [hep-th/9611134].


%%%


%\cite{Hull:2003kr}
\bibitem{Hull:2003kr}
C.~M.~Hull and A.~Catal-Ozer, ``Compactifications with S-duality
twists,'' JHEP {\bf 0310}, 034 (2003) [arXiv:hep-th/0308133].
%%CITATION = HEP-TH 0308133;%%


%\cite{Dall'Agata:2005ff}
\bibitem{Dall'Agata:2005ff}
  G.~Dall'Agata and S.~Ferrara,
  ``Gauged supergravity algebras from twisted tori compactifications with
  fluxes,''
  arXiv:hep-th/0502066.
  %%CITATION = HEP-TH 0502066;%%


%\cite{Hull:1998vy}
\bibitem{Hull:1998vy}
C.~M.~Hull, ``Massive string theories from M-theory and
F-theory,'' JHEP {\bf 9811}, 027 (1998) [arXiv:hep-th/9811021].
%%CITATION = HEP-TH 9811021;%%

%\cite{Kachru:2002sk}
\bibitem{Kachru:2002sk}
S.~Kachru, M.~B.~Schulz, P.~K.~Tripathy and S.~P.~Trivedi, ``New
supersymmetric string compactifications,'' JHEP {\bf 0303}, 061
(2003) [arXiv:hep-th/0211182].
%%CITATION = HEP-TH 0211182;%%


%\cite{deBoer:2001px}
\bibitem{deBoer:2001px}
J.~de Boer, R.~Dijkgraaf, K.~Hori, A.~Keurentjes, J.~Morgan,
D.~R.~Morrison and S.~Sethi, ``Triples, fluxes, and strings,''
Adv.\ Theor.\ Math.\ Phys.\  {\bf 4}, 995 (2002)
[arXiv:hep-th/0103170].
%%CITATION = HEP-TH 0103170;%%

%\cite{Kachru:2002he}
\bibitem{Kachru:2002he}
S.~Kachru, M.~B.~Schulz and S.~Trivedi, ``Moduli stabilization
from fluxes in a simple IIB orientifold,'' JHEP {\bf 0310}, 007
(2003) [arXiv:hep-th/0201028].
%%CITATION = HEP-TH 0201028;%%



%\cite{Hull:1994ys}
\bibitem{Hull:1994ys}
C.~M.~Hull and P.~K.~Townsend, ``Unity of superstring dualities,''
Nucl.\ Phys.\ B {\bf 438}, 109 (1995) [arXiv:hep-th/9410167].
%%CITATION = HEP-TH 9410167;%%


%\cite{Flournoy:2004vn}
\bibitem{Flournoy:2004vn}
A.~Flournoy, B.~Wecht and B.~Williams, ``Constructing nongeometric
vacua in string theory,'' Nucl.\ Phys.\ B {\bf 706}, 127 (2005)
[arXiv:hep-th/0404217].
%%CITATION = HEP-TH 0404217;%%


%\cite{Hull:2004in}
\bibitem{Hull:2004in}
C.~M.~Hull, ``A geometry for non-geometric string backgrounds,''
arXiv:hep-th/0406102.
%%CITATION = HEP-TH 0406102;%%


%\cite{Hellerman:2002ax}
\bibitem{Hellerman:2002ax}
S.~Hellerman, J.~McGreevy and B.~Williams, ``Geometric
constructions of nongeometric string theories,'' JHEP {\bf 0401},
024 (2004) [arXiv:hep-th/0208174].
%%CITATION = HEP-TH 0208174;%%


%\cite{Figueroa-O'Farrill:1994ns}
\bibitem{Figueroa-O'Farrill:1994ns}
J.~M.~Figueroa-O'Farrill and S.~Stanciu, ``Equivariant cohomology
and gauged bosonic sigma models,'' arXiv:hep-th/9407149.
%%CITATION = HEP-TH 9407149;%%

%\cite{Meessen:1998qm}
\bibitem{Meessen:1998qm}
  P.~Meessen and T.~Ortin,
  ``An Sl(2,Z) multiplet of nine-dimensional type II supergravity theories,''
  Nucl.\ Phys.\ B {\bf 541} (1999) 195
  [arXiv:hep-th/9806120].
  %%CITATION = HEP-TH 9806120;%%



%\cite{Vafa:1996xn}
\bibitem{Vafa:1996xn}
  C.~Vafa,
   ``Evidence for F-Theory,''
  %
  Nucl.\ Phys.\ B {\bf 469}, 403 (1996)
  [arXiv:hep-th/9602022].
  %%CITATION = HEP-TH 9602022;%%

  %\cite{Kumar:1996zx}
\bibitem{Kumar:1996zx}
  A.~Kumar and C.~Vafa,
   ``U-manifolds,''
  %
  Phys.\ Lett.\ B {\bf 396}, 85 (1997)
  [arXiv:hep-th/9611007].
  %%CITATION = HEP-TH 9611007;%%

%\cite{DeWolfe:1998eu}
\bibitem{DeWolfe:1998eu}
  O.~DeWolfe, T.~Hauer, A.~Iqbal and B.~Zwiebach,
  ``Uncovering the symmetries on (p,q) 7-branes: Beyond the Kodaira
  classification,''
  Adv.\ Theor.\ Math.\ Phys.\  {\bf 3} (1999) 1785
  [arXiv:hep-th/9812028].
  %%CITATION = HEP-TH 9812028;%%



%\cite{Bergshoeff:2003ri}
\bibitem{Bergshoeff:2003ri}
  E.~Bergshoeff, U.~Gran, R.~Linares, M.~Nielsen, T.~Ortin and D.~Roest,
  ``The Bianchi classification of maximal D = 8 gauged supergravities,''
  Class.\ Quant.\ Grav.\  {\bf 20} (2003) 3997
  [arXiv:hep-th/0306179].
  %%CITATION = HEP-TH 0306179;%%


%\cite{Maharana:my}
\bibitem{Maharana:my}
J.~Maharana and J.~H.~Schwarz, ``Noncompact Symmetries In String
Theory,'' Nucl.\ Phys.\ B {\bf 390}, 3 (1993)
[arXiv:hep-th/9207016].
%%CITATION = HEP-TH 9207016;%%

%\cite{Giveon:1988tt}
\bibitem{Giveon:1988tt}
  A.~Giveon, E.~Rabinovici and G.~Veneziano,
  ``Duality In String Background Space,''
  Nucl.\ Phys.\ B {\bf 322} (1989) 167.
  %%CITATION = NUPHA,B322,167;%%


%\cite{M-Theory paper}
\bibitem{M-Theory paper}
C.~Hull and R.A.~Reid-Edwards, ``Flux Compactifications of
M-Theory on Twisted Tori'' To appear shortly.


%\cite{Hull:1983yf}
\bibitem{Hull:1983yf}
  C.~M.~Hull,
  ``The Construction Of New Gauged N=8 Supergravities,''
  Physica {\bf 15D} (1985) 230.
  %%CITATION = PHYSA,15D,230;%%


%\cite{Hull:1984yy}
\bibitem{Hull:1984yy}
  C.~M.~Hull,
  ``A New Gauging Of N=8 Supergravity,''
  Phys.\ Rev.\ D {\bf 30} (1984) 760.
  %%CITATION = PHRVA,D30,760;%%

%\cite{Hull:1984qz}
\bibitem{Hull:1984qz}
  C.~M.~Hull,
  ``More Gaugings Of N=8 Supergravity,''
  Phys.\ Lett.\ B {\bf 148} (1984) 297.
  %%CITATION = PHLTA,B148,297;%%





%\cite{Lowe:2003qy}
\bibitem{Lowe:2003qy}
D.~A.~Lowe, H.~Nastase and S.~Ramgoolam, ``Massive IIA string
theory and matrix theory compactification,'' Nucl.\ Phys.\ B {\bf
667} (2003) 55 [arXiv:hep-th/0303173].
%%CITATION = HEP-TH 0303173;%%




\end{thebibliography}
\end{document}